%% file: paper.tex
\RequirePackage{fix-cm}
\documentclass{article}
\pdfoutput=1

\usepackage{microtype}
\usepackage{graphicx}
\usepackage{subfigure}
\usepackage{booktabs} %
\usepackage{pifont} %
\usepackage[dvipsnames]{xcolor}
\usepackage{listings}
\usepackage{mdframed}
\usepackage{xspace}
\usepackage{xurl}
\usepackage{tikz}
\usetikzlibrary{arrows.meta}
\usepackage{pgfplots}
\usepackage{soul}
\pgfplotsset{compat=1.18}
\usepgfplotslibrary{groupplots}
\input{figures/palette}
\input{figures/e2e-style}

\usepackage{hyperref}

\usepackage[preprint]{mlsys2025}

\mlsystitlerunning{HiPerfCC-KVCache: Accelerating Confidential LLM Serving with KV Caching}

\begin{document}
\input{macro}

\twocolumn[
\mlsystitle{\sysname: Accelerating Confidential LLM Serving with Hierarchical KV Caching}

\begin{mlsysauthorlist}
\mlsysauthor{Haoxuan Xu}{usc}
\mlsysauthor{Shaoyu Wang}{usc}
\mlsysauthor{Minghao Lin}{usc}
\mlsysauthor{Beijie Liu}{usc}
\mlsysauthor{Potung Yu}{usc}
\mlsysauthor{Yongqin Wang}{usc}
\mlsysauthor{Vincent Quentin Ulitzsch}{mit}
\mlsysauthor{Seo Jin Park}{usc}
\mlsysauthor{Murali Annavaram}{usc}
\mlsysauthor{Mengyuan Li}{usc}
\end{mlsysauthorlist}

\mlsysaffiliation{usc}{University of Southern California}
\mlsysaffiliation{mit}{Massachusetts Institute of Technology}

\mlsyskeywords{GPU confidential computing, Hierarchical KV caching, LLM serving, Driver contention, Transfer scheduling}

\vskip 0.3in

\begin{abstract}
Hierarchical key--value (KV) caching reduces repeated prefill
computation in large language model serving by retaining
reusable state beyond GPU memory.
Under GPU confidential computing (CC), however, protected
transfers can erase this benefit.
We show that bulk KV copies hold driver locks during prolonged
CPU-side processing, stalling independent serving operations
despite separate threads and streams.
Our characterization on Hopper and Blackwell GPUs shows how
lock scope, copy granularity, and allocation ownership affect
foreground progress and transfer parallelism.
Guided by these findings, we present \sysname, which manages
resource lifetimes, prioritizes ready token readback between
bounded backup submissions, and parallelizes KV restores
through auxiliary contexts with shared GPU staging.
Implemented in SGLang, \sysname uses existing protected
transfer interfaces without hardware or driver changes.
Across three models and three workloads on H200, \sysname
achieves a geometric-mean output-token throughput speedup
of $1.74\times$ (up to $2.72\times$) over the staged
KV-transfer \baseline baseline at the highest tested loads.
Mean time to first token decreases by 36\% on average
and up to 59\%.
The benefits generalize to RTX PRO 6000 Blackwell,
with a geometric-mean throughput speedup of $1.28\times$
(up to $2.20\times$) over the same baseline.
A capacity analysis further models and explains variation in gains
and remaining gaps to CC-off serving.
\end{abstract}
]

\printAffiliationsAndNotice{}

\input{sections/01-introduction}
\input{sections/02-background}
\input{sections/03-motivation-revised}
\input{sections/04-characterization-revised}
\input{sections/05-design-revised}
\input{sections/06-evaluation}

\input{sections/07-related-work}
\input{sections/09-conclusion}

\bibliography{ref}
\bibliographystyle{mlsys2025}

\section*{AI Usage Disclosure}
Claude Code and Codex were used under the authors' supervision to assist with
language polishing of the manuscript, figure preparation, and portions of the
code implementation.
The authors reviewed and edited the AI-assisted text and code and take full
responsibility for the paper and code.

\clearpage
\appendix
\input{sections/10-appendix}

\end{document}

%% file: figures/palette.tex
\definecolor{VeloCCProtected}{HTML}{4E79A7}
\definecolor{VeloCCCipher}{HTML}{F28E2B}
\definecolor{VeloCCHighlight}{HTML}{E15759}
\definecolor{VeloCCInk}{HTML}{333333}
\definecolor{VeloCCBackup}{HTML}{59A14F}
\definecolor{VeloCCToken}{HTML}{7B68B5}

\ifdefined\pdfmapfile\pdfmapfile{+libertine.map}\fi
\newcommand{\figurefont}{\fontencoding{T1}\fontfamily{LinuxBiolinumT-TLF}}

%% file: figures/e2e-style.tex
\usetikzlibrary{arrows.meta}
\pgfdeclareplotmark{filledstar}{%
  \pgfpathmoveto{\pgfqpointpolar{90}{\pgfplotmarksize}}%
  \pgfpathlineto{\pgfqpointpolar{54}{.382\pgfplotmarksize}}%
  \pgfpathlineto{\pgfqpointpolar{18}{\pgfplotmarksize}}%
  \pgfpathlineto{\pgfqpointpolar{-18}{.382\pgfplotmarksize}}%
  \pgfpathlineto{\pgfqpointpolar{-54}{\pgfplotmarksize}}%
  \pgfpathlineto{\pgfqpointpolar{-90}{.382\pgfplotmarksize}}%
  \pgfpathlineto{\pgfqpointpolar{-126}{\pgfplotmarksize}}%
  \pgfpathlineto{\pgfqpointpolar{-162}{.382\pgfplotmarksize}}%
  \pgfpathlineto{\pgfqpointpolar{-198}{\pgfplotmarksize}}%
  \pgfpathlineto{\pgfqpointpolar{-234}{.382\pgfplotmarksize}}%
  \pgfpathclose\pgfusepathqfillstroke
}
\pgfplotsset{
  e2e gpu/.style={VeloCCToken,solid,line width=.65pt,mark=triangle*,mark size=1pt,
    mark options={solid,fill=VeloCCToken,line width=.3pt}},
  e2e background/.style={VeloCCCipher,solid,line width=.65pt,mark=square*,mark size=.85pt,
    mark options={solid,fill=VeloCCCipher,line width=.25pt}},
  e2e velocc/.style={VeloCCProtected,solid,line width=.85pt,mark=filledstar,mark size=1.3pt,
    mark options={solid,fill=VeloCCProtected,line width=.25pt}},
  e2e cc off/.style={VeloCCBackup,solid,line width=.65pt,mark=*,mark size=1pt,
    mark options={solid,fill=VeloCCBackup,line width=.3pt}},
  e2e gpu legend/.style={e2e gpu,mark size=2.25pt},
  e2e background legend/.style={e2e background,mark size=1.8pt},
  e2e velocc legend/.style={e2e velocc,mark size=2.9pt},
  e2e cc off legend/.style={e2e cc off,mark size=2pt},
}

%% file: macro.tex
\newcommand{\sysname}{\texorpdfstring{\textulc{\textit{HiPerfCC-KVCache}}}{HiPerfCC-KVCache}\xspace}
\newcommand{\baseline}{\texorpdfstring{\textulc{HiCache-Workers}}{HiCache-Workers}\xspace}
\newcommand{\baselineshort}{HW\xspace}
\newcommand{\laegis}{L\AE{}GIS\xspace}
\newcommand{\bheading}[1]{{\vspace{2pt}\noindent{\textbf{#1}}}}
\newcommand{\secref}[1]{\mbox{Section~\ref{#1}}\xspace}
\newcommand{\figref}[1]{\mbox{Figure~\ref{#1}}}
\newcommand{\tabref}[1]{\mbox{Table~\ref{#1}}}
\newcommand{\appref}[1]{\mbox{Appendix~\ref{#1}}}

\newmdenv[
  backgroundcolor=VeloCCProtected!12,
  hidealllines=true,
  roundcorner=0pt,
  innerleftmargin=7pt,
  innerrightmargin=7pt,
  innertopmargin=5pt,
  innerbottommargin=5pt,
  skipabove=7pt,
  skipbelow=7pt,
  nobreak=true
]{takeawaybox}
\newcommand{\takeaway}[1]{%
  \begin{takeawaybox}%
  \noindent\textbf{Takeaway.} #1%
  \end{takeawaybox}%
}

\newenvironment{packeditemize}{
\begin{list}{$\bullet$}{
\setlength{\labelwidth}{0pt}
\setlength{\itemsep}{2pt}
\setlength{\leftmargin}{\labelwidth}
\addtolength{\leftmargin}{\labelsep}
\setlength{\parindent}{0pt}
\setlength{\listparindent}{\parindent}
\setlength{\parsep}{1pt}
\setlength{\topsep}{1pt}}}{\end{list}}

%% file: sections/01-introduction.tex
\section{Introduction}
\label{sec:intro}

Hosting large language model (LLM) services raises concerns about exposing sensitive inputs to privileged software and administrators~\cite{cloudcc,pipellm}.
Confidential computing (CC) addresses these concerns through hardware-enforced trusted execution environments (TEEs).
Current confidential GPU platforms combine CPU and GPU TEEs to protect the serving runtime, host memory, and GPU execution~\cite{gpucc}.
Prior studies report low overhead when computation remains within protected GPU memory~\cite{cloudcc,serializedbridge,blackwell,nvidiaccperf}.
However, workloads involving frequent CPU--GPU transfers can incur substantial encryption and authentication overhead.

This challenge is particularly important for \emph{hierarchical key--value (KV) caching}, a widely used optimization in LLM serving.
KV caching avoids redundant computation across requests sharing prompt prefixes or conversation histories~\cite{pagedattention,sglang,cachedattention,wild}.
As longer contexts and higher concurrency exceed GPU memory capacity, hierarchical caching extends KV storage into host memory, allowing previously computed KV states to be restored instead of recomputed~\cite{cachedattention,lmcache}.
This trades expensive prefill computation for CPU--GPU transfers, improving serving performance.
Under GPU CC, however, these transfers introduce costly software-encrypted bounce buffers in shared host memory~\cite{ccoverview,gpucc}, potentially outweighing the benefits of KV reuse.

Our measurements with SGLang demonstrate this performance reversal.
On a multi-turn workload serving Qwen2.5-14B, HiCache's two stock
backends reduce mean time to first token (TTFT) by 64--69\%
relative to GPU-only prefix caching with CC disabled (CC-off).
With CC enabled (CC-on), however, their mean TTFT is
$2.0$--$5.3\times$ slower that of the CC-on GPU-only baseline
(\secref{sec:motivation}).
This raises a fundamental question:
\textit{how can LLM serving retain the benefits of hierarchical KV caching under GPU CC?}

Existing approaches mitigate protected-transfer overhead through pre-encryption, transfer prediction, and runtime scheduling.
Architectural proposals such as \laegis enable pre-encryption but require GPU hardware changes~\cite{laegis}.
PipeLLM overlaps encryption with computation through transfer prediction~\cite{pipellm}, whereas KV-cache restores depend on future request arrivals and GPU-cache residency, making them difficult to predict.
Other work reduces transfer contention through synchronous scheduling and recovers overlap by offloading token readback to background workers~\cite{serializedbridge,blackwell}.
To the best of our knowledge, no prior paper has systematically optimized hierarchical KV caching under GPU CC.
To understand whether existing techniques are sufficient, we construct \emph{\baseline}, a strong baseline that consolidates the applicable software-level optimizations from prior work.
Specifically, it combines GPU staging, dedicated CPU workers and CUDA streams for KV backups and restores, and scheduling that defers requests awaiting restoration while continuing to serve ready requests.
Together, these optimizations reduce mean TTFT by 60\% relative to SGLang's vanilla HiCache-direct backend under CC, from 3.39~s to 1.35~s.
Nevertheless, despite achieving the highest throughput and lowest mean TTFT among the CC-on SGLang configurations we evaluate, \baseline reaches only 77\% of CC-off HiCache's throughput and still incurs $2.6\times$ the mean TTFT compared to HiCache when CC is disabled (\secref{sec:motivation}).

To understand this remaining gap, we trace protected KV transfers and foreground serving operations on Hopper and Blackwell GPUs.
We find that bulk protected copies perform prolonged CPU-side processing while holding driver locks shared with resource management, stream operations, and latency-sensitive foreground transfers (\secref{sec:characterization}).
Consequently, background copies can stall unrelated requests even across separate threads and CUDA streams.
Our analysis further identifies copy granularity and allocation ownership as important factors shaping transfer interference.
Thus, improving transfer bandwidth or moving copies to background workers alone is insufficient; efficient hierarchical caching must also minimize interference with foreground serving.

Guided by these findings, we present \sysname, a contention-aware runtime for hierarchical KV caching under GPU CC. Rather than treating protected transfers as a bandwidth problem alone, \sysname organizes them around the identified driver-lock structure so that background KV movement neither blocks ready requests nor serializes on itself. It does so through three mechanisms, each targeting one source of contention.
Firstly, \sysname reuses transfer resources and coordinates their lifetimes across backup and restore paths to reduce scheduler-side resource-management stalls.
Secondly, it bounds KV backup submissions and prioritizes latency-sensitive token readback to limit foreground interference while preserving transfer--computation overlap.
Thirdly, it uses auxiliary CUDA contexts to reduce contention and parallelize KV restoration, extending context-level parallelism previously explored for model loading~\cite{serializedbridge}.
By coordinating allocation ownership and sharing GPU staging buffers across contexts, \sysname also eliminates an additional assembly copy during restoration.
Together, these mechanisms address both protected-transfer overhead and its interference with ongoing serving.

We implement \sysname in SGLang using existing protected-transfer interfaces and evaluate three models on three workloads across H200 and RTX PRO 6000 Blackwell. At the highest tested loads on H200, \sysname achieves a geometric-mean throughput speedup of $1.74\times$ (up to $2.72\times$) over HiCache-Workers and reduces mean TTFT by 36\% on average. An approximate capacity model separates compute demand (including extra prefill and launch costs) from restoration demand (copy service and admission waits) to explain the gains and remaining gaps to CC-off. On SWE-bench, longer generation allows restoration to overlap computation without limiting throughput (the restore-unsaturated regime), and \sysname reaches 91--102\% of CC-off throughput. NarrativeQA's high restoration demand keeps it restore-saturated:
\sysname achieves $2.31$--$2.72\times$ throughput gains over
HiCache-Workers but reaches only 65--73\% of CC-off throughput. The same distinction explains results on Blackwell, where \sysname achieves a geometric-mean throughput speedup of $1.28\times$ over HiCache-Workers.

This paper makes the following contributions:
\begin{packeditemize}
    \item \textbf{HiCache overhead diagnosis under CC.}
    We reveal how protected KV transfers block unrelated serving work through driver-lock contention, and identify copy granularity and allocation ownership as additional constraints across Hopper and Blackwell.

    \item \textbf{Contention-aware KV backup.}
    We coordinate resource lifetimes and bound backup submissions while prioritizing latency-sensitive token readback.

    \item \textbf{Context-parallel KV restoration.}
    We combine auxiliary CUDA contexts with cross-context buffer sharing to parallelize restores and avoid an additional assembly copy.

    \item \textbf{End-to-end gains and capacity analysis.}
    \sysname achieves geometric-mean throughput speedups of
    $1.74\times$ on H200 and $1.28\times$ on Blackwell over
    HiCache-Workers, and reduces mean TTFT on H200 by 36\%
    on average. Ablations and capacity modeling explain
    these gains and the remaining gaps to CC-off.

\end{packeditemize}

Our code is available at\\
\href{https://github.com/sept-usc/HiPerfCC-KVCache}{\mbox{\texttt{github.com/sept-usc/HiPerfCC-KVCache}}}.

\bheading{Scope of the work.}
\sysname targets single-GPU LLM serving on existing GPU CC
platforms that use software-encrypted bounce buffers for
CPU--GPU transfers, including NVIDIA Hopper and
Blackwell~\cite{nvidia_secure_ai_2025}.
Its software-only design enables deployment without hardware
or driver modifications.
While future hardware-assisted TEE-I/O mechanisms such as PCIe
Integrity and Data Encryption (IDE) and TEE Device Interface
Security Protocol (TDISP) can eliminate bounce buffers and
support fast multi-GPU serving, they require additional
end-to-end hardware and software
support~\cite{gpu_travelling_ccs25,nvidia_secure_ai_2025,pcisig_tdisp_2022}.
We therefore focus on existing bounce-buffer-based CC
deployments, leaving TEE-I/O and multi-GPU serving outside
our scope.

%% file: sections/02-background.tex
\section{Background}
\label{sec:bg}

\subsection{Hierarchical KV Caching}
\label{sec:bg:kv}

Serving systems retain KV blocks for reuse across requests and
conversation turns~\cite{pagedattention,sglang,cachedattention}.
Hierarchical caches extend reuse beyond GPU capacity through
host memory and other storage tiers~\cite{cachedattention,lmcache}.
We focus on the host tier: \emph{backup} copies KV from GPU to
host (D2H), and \emph{restore} brings it back (H2D).
Since blocks may be scattered across the GPU KV pool, a kernel
gathers them into GPU staging for bulk D2H copying. Restoration
performs H2D copying followed by scatter~\cite{lmcache}.

Both paths can overlap independent computation~\cite{cachedattention,lmcache}.
Unless stated otherwise, we use \emph{write-through} backups,
initiated when blocks are cached on the GPU~\cite{hicache}.
Once backed up and no longer in use, blocks can be reclaimed
without another D2H copy.
For restores, the scheduler can overlap transfers with computation
by executing ready requests while deferring those awaiting KV~\cite{vllmoffload}.

\subsection{GPU Confidential Computing}
\label{sec:bg:cc}

NVIDIA GPU confidential computing combines a confidential virtual
machine (CVM), which protects the serving runtime and private host
memory, with a GPU TEE that protects device execution and memory~\cite{ccoverview,gpucc}.

On the protected CPU--GPU transfer path studied here, the GPU
cannot directly DMA into CVM private memory.
Transfers therefore pass through \emph{bounce buffers} in shared
host memory outside the CVM's private-memory protection, carrying
authenticated ciphertext (\figref{fig:cc-transfer}).
For H2D, the driver encrypts and authenticates data on the CPU
using AES-GCM, and the GPU copy engine verifies and decrypts it.
D2H reverses these roles~\cite{ccoverview,gpucc,pipellm}.

\input{figures/cc-transfer}

\bheading{Threat Model.}
We inherit the GPU CC platform's threat model~\cite{gpucc},
trusting the CPU/GPU TEEs and software inside the CVM,
while treating the host OS, hypervisor, and CPU--GPU
interconnect as untrusted.
\sysname uses existing protected-transfer interfaces
without changing these trust assumptions.

\subsection{CUDA Execution and Host Memory}
\label{sec:bg:cuda}

A CUDA \emph{context} manages device resources and execution state.
Within a context, \emph{streams} order submitted copies and kernels,
while \emph{events} track completion and express dependencies
between streams~\cite{cudaguide}.
Independent operations in different streams may execute concurrently.

Host buffers used by these operations can be allocated through
different interfaces.
\emph{CUDA host allocations}, obtained through interfaces such as
\texttt{cudaHostAlloc}, conventionally provide page-locked (pinned)
memory, while \emph{ordinary host memory} allocated through
\texttt{malloc} is pageable~\cite{cudaguide}.
These application buffers are distinct from the shared host bounce
buffers carrying ciphertext.

Host-memory type and transfer direction affect the host-side
behavior of asynchronous copy APIs such as
\texttt{cudaMemcpyAsync}~\cite{cudaguide}.
Although these calls may return before the device transfer completes,
CPU-side encryption or decryption under CC can keep the calling
thread occupied~\cite{pipellm,serializedbridge}.
Submitting the copy to another stream therefore does not, by itself,
move this CPU-side work off the calling thread.

%% file: figures/cc-transfer.tex
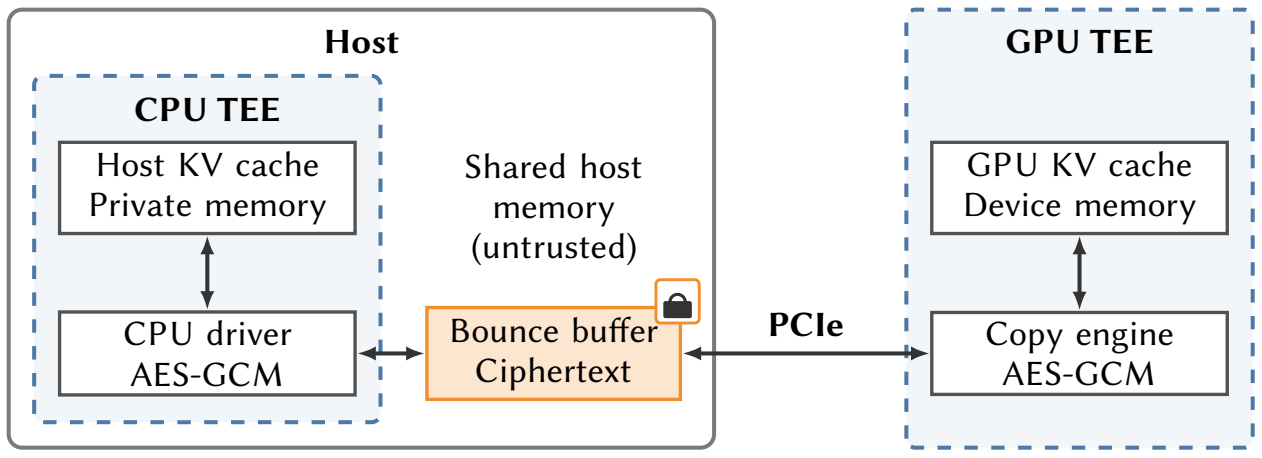
\begin{figure}[t]
\centering
\resizebox{\columnwidth}{!}{%
\begin{tikzpicture}[
    font=\figurefont\fontsize{8}{9}\selectfont,
    heading/.style={font=\figurefont\bfseries\fontsize{8}{9}\selectfont},
    box/.style={draw=VeloCCInk!85, line width=.75pt, fill=white, align=center,
                text width=2.18cm, minimum height=.57cm, inner sep=2pt},
    flow/.style={{Latex[length=1.5mm]}-{Latex[length=1.5mm]}, line width=.9pt, draw=VeloCCInk}]
    \draw[draw=VeloCCInk!65, line width=.9pt, rounded corners=3pt]
        (-1.56,-.42) rectangle (3.98,3.02);
    \node[heading] at (1.21,2.77) {Host};
    \draw[dashed, line width=.9pt, rounded corners=2pt, draw=VeloCCProtected, fill=VeloCCProtected!7]
        (-1.36,-.22) rectangle (1.36,2.50);
    \draw[dashed, line width=.9pt, rounded corners=2pt, draw=VeloCCProtected, fill=VeloCCProtected!7]
        (5.49,-.42) rectangle (8.21,3.02);
    \node[heading] at (0,2.24) {CPU TEE};
    \node[heading] at (6.85,2.77) {GPU TEE};
    \node[box] (host) at (0,1.62) {Host KV cache\\Private memory};
    \node[box] (cpu) at (0,.32) {CPU driver\\AES-GCM};
    \node[box, text width=1.85cm, draw=VeloCCCipher, fill=VeloCCCipher!22] (bounce)
        at (2.72,.32) {Bounce buffer\\Ciphertext};
    \begin{scope}[shift={([xshift=-1pt,yshift=1pt]bounce.north east)}]
        \draw[rounded corners=1pt, draw=VeloCCCipher, fill=white, line width=.65pt]
            (-.17,-.17) rectangle (.17,.17);
        \draw[draw=VeloCCInk, line width=.7pt]
            (-.06,-.005) arc (180:0:.06);
        \draw[draw=VeloCCInk, fill=VeloCCInk, line width=.4pt, rounded corners=.3pt]
            (-.105,-.005) rectangle (.105,-.125);
    \end{scope}
    \node[box] (gpu) at (6.85,.32) {Copy engine\\AES-GCM};
    \node[box] (device) at (6.85,1.62) {GPU KV cache\\Device memory};
    \draw[flow] (host.south) -- (cpu.north);
    \draw[flow] (cpu) -- (bounce);
    \draw[flow] (bounce) -- node[above, heading] {PCIe} (gpu);
    \draw[flow] (gpu.north) -- (device.south);
    \node[align=center, text width=1.85cm] at (2.72,1.45)
        {Shared host\\memory\\(untrusted)};
\end{tikzpicture}%
}
\caption{KV movement between CPU and GPU TEEs. Payloads outside the TEEs are software-encrypted and authenticated.}
\label{fig:cc-transfer}
\end{figure}

%% file: sections/03-motivation-revised.tex
\section{Limits of Existing Approaches}
\label{sec:motivation}

Host KV caching avoids repeated prefill computation, but protected
transfers can offset these savings under GPU CC.
We examine how existing transfer configurations and runtime
optimizations recover this benefit, focusing on output-token throughput
and mean time to first token (TTFT) relative to CC-off host caching.

We run Qwen2.5-14B on a single H200 GPU using a multi-turn workload.
The reusable KV state exceeds GPU memory and fits in the host memory.
By default, CPU-side batch preparation and output processing overlap
with GPU execution.
Full settings appear in \appref{sec:appendix:motivation}.
\figref{fig:motivation-ttft} shows throughput and TTFT across SGLang
configurations, with CC-off setting.

\input{figures/motivation-ttft}

\bheading{Transfer configuration.}
We use SGLang with HiCache~\cite{hicache} as our primary platform.
HiCache provides a kernel backend that accesses host KV through
GPU kernels and a direct backend that uses explicit
\texttt{cudaMemcpyAsync} calls~\cite{hicachedesign}.
With CC disabled, the kernel backend achieves lower mean TTFT
than the direct backend and serves as our host-cache reference.
Under CC, both stock backends lose their TTFT advantage over
GPU-only serving, although the direct backend performs better
than the kernel backend.

We therefore apply GPU staging (\secref{sec:bg:kv}) to the
direct backend, consolidating its many small KV copies into
larger transfers for both backup and restore.
The resulting \emph{HiCache-staged} roughly halves mean TTFT
relative to the CC-on direct backend, but reaches only 72\% of
the CC-off HiCache-kernel reference's throughput and incurs
$3.2\times$ its mean TTFT. \tabref{tab:motivation-stock-backends} in
\appref{sec:appendix:motivation} reports detailed results.

\bheading{Speculative encryption with PipeLLM.}
PipeLLM exploits regular memory-swapping patterns to predict
upcoming H2D transfers and overlap their encryption with GPU
computation~\cite{pipellm}.
For prefix caching, restore order depends on future request
arrivals and GPU-cache residency, so backup order alone does not
determine which KV to restore next.
We integrate PipeLLM with HiCache-staged in SGLang.
On our workload, only 0.016\% of restored KV bytes use
pre-encrypted ciphertext, while throughput decreases and mean
TTFT increases relative to HiCache-staged.
With such limited coverage, the added speculative work can
outweigh the savings on the restore path.

\bheading{Synchronous scheduling.}
Following Serialized Bridge~\cite{serializedbridge}, we disable
overlap scheduling in HiCache-staged, yielding \emph{HiCache-Sync}.
The scheduler waits for GPU execution and output readback
before processing results and preparing the next batch.
This avoids deferred output processing but sacrifices CPU--GPU
overlap across batches.
Relative to HiCache-staged, mean TTFT decreases by about 13\%,
but throughput falls by 1.6\%.

\bheading{KV-transfer workers.}
\label{sec:motivation:strawman}
Following prior work~\cite{serializedbridge,blackwell},
our overlap-scheduled configurations offload blocking
token-readback calls to a CPU response worker.
We extend this approach to KV backup and restore, constructing
\emph{\baseline} with separate CPU workers and CUDA streams
within the serving context.
It retains HiCache-staged's GPU staging and overlap scheduling.
The scheduler defers requests awaiting restored KV so ready
requests can proceed.

Relative to HiCache-staged, \baseline increases throughput by 6.7\%
and reduces mean TTFT by about 18\%, yielding the highest throughput
and lowest mean TTFT among the evaluated CC-on SGLang configurations.
We therefore use it as our baseline for subsequent analysis
and evaluation.
However, it reaches only 77\% of the CC-off HiCache-kernel
reference's throughput, with $2.6\times$ its mean TTFT.

%% file: figures/motivation-ttft.tex
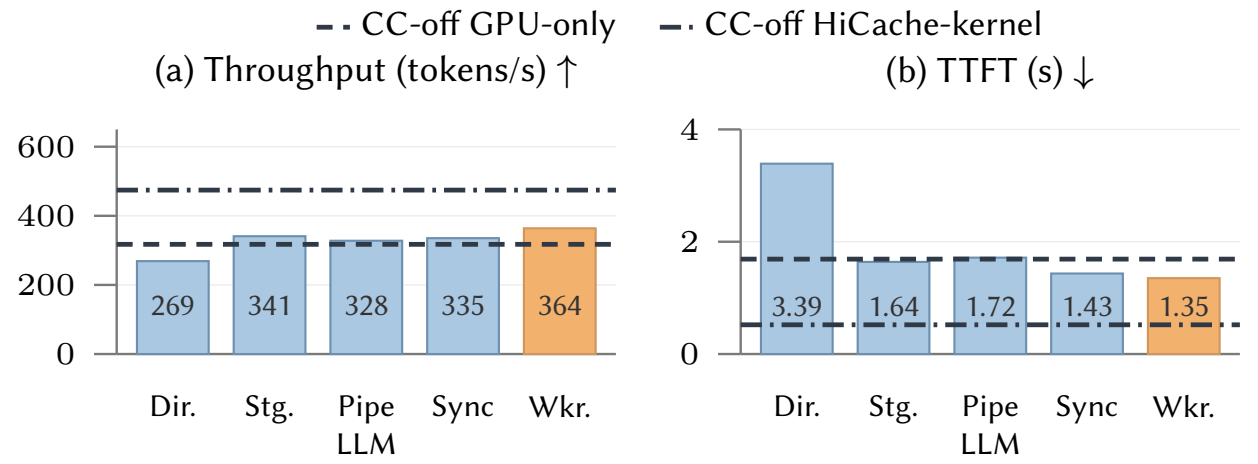
\begin{figure}[!t]
\centering
\input{figures/motivation-ttft-plot}\par
\caption{Output-token throughput and mean TTFT in SGLang.
Bars: CC-on; lines: CC-off.
Dir./Stg./Wkr.: HiCache-direct/staged/Workers.}
\label{fig:motivation-ttft}
\end{figure}

%% file: figures/motivation-ttft-plot.tex
\begingroup%
\def\motDirect{3.389760962498895}%
\def\motDirectThroughput{268.847134767273985}%
\def\motDirectLabel{3.39}%
\def\motDirectThroughputLabel{269}%
\def\motStaged{1.640415313115523}%
\def\motStagedThroughput{340.956994427505492}%
\def\motStagedLabel{1.64}%
\def\motStagedThroughputLabel{341}%
\def\motPipeLLM{1.716597258793695}%
\def\motPipeLLMThroughput{327.883089884851188}%
\def\motPipeLLMLabel{1.72}%
\def\motPipeLLMThroughputLabel{328}%
\def\motSync{1.433757890318850}%
\def\motSyncThroughput{335.345782500520556}%
\def\motSyncLabel{1.43}%
\def\motSyncThroughputLabel{335}%
\def\motWorkers{1.351534875265736}%
\def\motWorkersThroughput{363.822913053383218}%
\def\motWorkersLabel{1.35}%
\def\motWorkersThroughputLabel{364}%
\def\motGPUOff{1.690247821712499}%
\def\motGPUOffThroughput{317.466190620521502}%
\def\motGPUOffLabel{1.69}%
\def\motGPUOffThroughputLabel{317}%
\def\motCacheOff{0.520727472846877}%
\def\motCacheOffThroughput{474.422730303716605}%
\def\motCacheOffLabel{0.52}%
\def\motCacheOffThroughputLabel{474}%
\definecolor{MotBar}{HTML}{ABC8E2}%
\definecolor{MotWorker}{HTML}{F3B16E}%
\definecolor{MotBarEdge}{HTML}{6C8FB0}%
\definecolor{MotReference}{HTML}{303946}%
\pgfplotsset{
  motivation axis/.style={scale only axis,
    width=.40\linewidth,height=.18\linewidth,
    xmin=-.58,xmax=4.58,xtick={0,1,2,3,4},
    xticklabels={{Dir.},{Stg.},{Pipe\\LLM},{Sync},{Wkr.}},
    xticklabel style={align=center,font=\figurefont\fontsize{6}{7}\selectfont},
    yticklabel style={font=\figurefont\fontsize{6}{7}\selectfont},
    title style={font=\figurefont\fontsize{7}{8}\selectfont,yshift=-2pt},
    xtick style={draw=none},axis lines*=left,
    axis line style={VeloCCInk!65,line width=.45pt},
    tick align=outside,tick style={VeloCCInk!65,line width=.45pt},
    ymajorgrids=true,grid style={VeloCCInk!10,line width=.25pt},
    scaled ticks=false,clip=false},
  motivation bar/.style={ybar,bar shift=0pt,bar width=13.5pt,
    line width=.4pt,fill=MotBar,draw=MotBarEdge,forget plot},
  motivation gpu ref/.style={MotReference,line width=.9pt,
    dash pattern=on 3pt off 2pt},
  motivation cache ref/.style={MotReference,line width=.9pt,
    dash pattern=on 4pt off 1.5pt on .7pt off 1.5pt}
}%
\begin{tikzpicture}
\begin{axis}[motivation axis,name=rate,
  ymin=0,ymax=650,ytick={0,200,400,600},
  title={(a) Throughput (tokens/s) $\uparrow$}]
\addplot[motivation bar] coordinates {(0,\motDirectThroughput)};
\addplot[motivation bar] coordinates {(1,\motStagedThroughput)};
\addplot[motivation bar] coordinates {(2,\motPipeLLMThroughput)};
\addplot[motivation bar] coordinates {(3,\motSyncThroughput)};
\addplot[motivation bar,fill=MotWorker,draw=MotWorker!85!black]
    coordinates {(4,\motWorkersThroughput)};
\addplot[motivation gpu ref]
    coordinates {(-.58,\motGPUOffThroughput) (4.58,\motGPUOffThroughput)};
\addplot[motivation cache ref]
    coordinates {(-.58,\motCacheOffThroughput) (4.58,\motCacheOffThroughput)};
\foreach \x/\value in {0/\motDirectThroughputLabel,1/\motStagedThroughputLabel,2/\motPipeLLMThroughputLabel,3/\motSyncThroughputLabel,4/\motWorkersThroughputLabel}{%
    \edef\labelpoint{\noexpand\node[anchor=center,inner sep=.4pt,text=VeloCCInk,
        font=\noexpand\figurefont\noexpand\fontsize{6}{8}\noexpand\selectfont]
        at (axis cs:\x,140)
        {\value};}
    \labelpoint
}
\end{axis}
\begin{axis}[motivation axis,name=lat,at={(rate.south east)},anchor=south west,
  xshift=.10\linewidth,ymin=0,ymax=4,ytick={0,2,4},
  title={(b) TTFT (s) $\downarrow$}]
\addplot[motivation bar] coordinates {(0,\motDirect)};
\addplot[motivation bar] coordinates {(1,\motStaged)};
\addplot[motivation bar] coordinates {(2,\motPipeLLM)};
\addplot[motivation bar] coordinates {(3,\motSync)};
\addplot[motivation bar,fill=MotWorker,draw=MotWorker!85!black]
    coordinates {(4,\motWorkers)};
\addplot[motivation gpu ref]
    coordinates {(-.58,\motGPUOff) (4.58,\motGPUOff)};
\addplot[motivation cache ref]
    coordinates {(-.58,\motCacheOff) (4.58,\motCacheOff)};
\foreach \x/\value in {0/\motDirectLabel,1/\motStagedLabel,2/\motPipeLLMLabel,3/\motSyncLabel,4/\motWorkersLabel}{%
    \edef\labelpoint{\noexpand\node[anchor=center,inner sep=.4pt,text=VeloCCInk,
        font=\noexpand\figurefont\noexpand\fontsize{6}{8}\noexpand\selectfont]
        at (axis cs:\x,.85)
        {\value};}
    \labelpoint
}
\end{axis}
\path (rate.north west)--coordinate[pos=.5] (legendmid) (lat.north east);
\node[anchor=south,inner sep=0pt,yshift=16pt,
    font=\figurefont\fontsize{6.7}{8}\selectfont] at (legendmid) {
  \tikz[baseline=-.5ex]\draw[MotReference,line width=.9pt,dash pattern=on 3pt off 2pt] (0,0)--(.22,0);
  CC-off GPU-only\hspace{6pt}
  \tikz[baseline=-.5ex]\draw[MotReference,line width=.9pt,dash pattern=on 4pt off 1.5pt on .7pt off 1.5pt] (0,0)--(.22,0);
  CC-off HiCache-kernel};
\end{tikzpicture}%
\endgroup%

%% file: sections/04-characterization-revised.tex
\section{Demystifying Stalls During KV Transfers}
\label{sec:characterization}
\label{sec:characterization:locks}

We examine why \baseline still leaves a substantial
TTFT gap under CC.
Our analysis connects serving stalls during KV transfers to
contention within the CUDA driver.

\bheading{Observed serving stalls.}
\figref{fig:baseline-timeline} summarizes two stall patterns
from serving traces of \baseline on H200 under CC.
In (a), the scheduler stalls during a D2H copy in the backup
worker, leaving the compute stream idle before the next prefill.
In (b), an H2D copy in the restore worker delays submission
of another request's prefill, even though that request's KV
already resides on the GPU.
These stalls interrupt ready work even though the KV copies
run on separate workers and streams.

\input{figures/baseline-timeline}

\bheading{Driver-lock contention.}
To identify what blocks the scheduler during these stalls,
we use targeted CUDA API probes and driver-lock tracing on
H200 and RTX PRO 6000 under CC.
Bulk \texttt{cudaMemcpyAsync} and \texttt{cudaMemcpyBatchAsync}
calls acquire three classes of user-space driver locks.
Both platforms exhibit the same lock scopes and holding patterns.
Details of the foreground API probes appear in \appref{sec:appendix:characterization}.

\emph{Context-wide resource management (C).}
The copy holds a context-wide read/write lock in read mode
($C_{\mathrm{R}}$) throughout host-side transfer processing.
Resource-management operations requiring write access
($C_{\mathrm{W}}$), such as event creation and destruction, wait
regardless of stream choice.

\emph{Stream submission (S).}
The copy exclusively holds its stream's submission mutex until
just before the host API call returns.
Kernel launches, event recording, and further copies on that stream
wait on S, while operations on another stream use a different instance.

\emph{Same-direction transfers within a context (T).}
Same-direction pinned copies share a context-wide mutex across streams. Opposite directions and contexts use different T instances. Single-buffer calls hold T throughout host-side processing. Batched calls release and reacquire T between buffers, but a waiting foreground copy can remain blocked across multiple acquisitions. In uninstrumented probes, most same-direction foreground copies on other streams within the context complete only after the batch returns.

\input{tables/cuda-api-locks}

\tabref{tab:cuda-api-locks} summarizes the lock acquisitions of
foreground operations.
Separate streams use different S instances but share C and
same-direction T.
With CC disabled, the same pinned-copy probes do not acquire T,
and release C and S shortly after transfer submission.
Under CC, CPU encryption or decryption extends the host calls
and their lock holds, increasing the waits of conflicting operations.

A diagnostic replay of \emph{\baseline} matches foreground lock
waits to overlapping bulk-copy lock holds in the stall patterns
of \figref{fig:baseline-timeline}.
During backups, the scheduler's event creation and destruction
require $C_{\mathrm{W}}$ and wait behind the backup's
$C_{\mathrm{R}}$ hold.
During restores, an 8-byte pinned metadata upload waits 73.7~ms
on T behind a 384~MiB H2D copy on another stream.
This delays batch launch even though the batch's KV already
resides on the GPU.

These lock scopes and holding patterns guide three runtime choices:
when to create and reclaim resources, how to bound and order copy
submissions, and which transfers share a context.
\secref{sec:design} applies these choices to write-through backups
and on-demand restores.

%% file: figures/baseline-timeline.tex
\begin{figure}[t]
\centering
\input{figures/baseline-timeline-plot}
\caption{Schematic serving stalls in \baseline under CC.
Backup and restore use separate workers, both labeled ``KV worker''.
Red marks batch submission.}
\label{fig:baseline-timeline}
\end{figure}
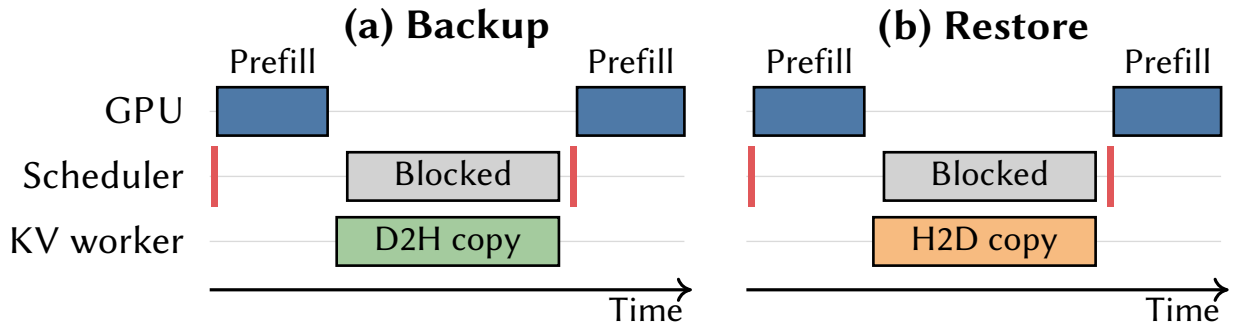

%% file: figures/baseline-timeline-plot.tex
\begingroup
\begin{tikzpicture}[
  x=1cm,y=1cm,
  font=\figurefont\fontsize{8}{9.2}\selectfont,
  lane/.style={anchor=mid east,inner sep=0pt,font=\figurefont\fontsize{7.5}{8.5}\selectfont},
  barlabel/.style={inner sep=0pt,text height=4.5pt,text depth=1.3pt,
    font=\figurefont\fontsize{6.8}{7.4}\selectfont},
  paneltitle/.style={font=\figurefont\bfseries\fontsize{8}{9}\selectfont,inner sep=0pt},
  gpu/.style={draw=black,line width=.5pt,fill=VeloCCProtected},
  blocked/.style={draw=black,line width=.5pt,fill=VeloCCInk!22},
  backup/.style={draw=black,line width=.5pt,fill=VeloCCBackup!55},
  restore/.style={draw=black,line width=.5pt,fill=VeloCCCipher!60},
  guide/.style={draw=VeloCCInk!18,line width=.25pt},
  submit/.style={fill=VeloCCHighlight,draw=none},
  axis/.style={-{To[length=1mm,width=1.6mm]},line width=.6pt},
]
\path[use as bounding box] (0,-1.42) rectangle (8.25,.64);
\node[lane] at (1.22,0) {GPU};
\node[lane] at (1.22,-.43) {Scheduler};
\node[lane] at (1.22,-.86) {KV worker};

\foreach \offset/\title/\copystyle/\copylabel in {
  1.38/{(a) Backup}/backup/{D2H copy},
  4.93/{(b) Restore}/restore/{H2D copy}
}{%
  \begin{scope}[xshift=\offset cm]
    \node[paneltitle] at (1.57,.55) {\title};
    \foreach \y in {0,-.43,-.86}
      \draw[guide] (0,\y) -- (3.14,\y);

    \draw[gpu] (.05,-.16) rectangle (.78,.16);
    \draw[gpu] (2.43,-.16) rectangle (3.14,.16);
    \node[barlabel,anchor=south] at (.415,.20) {Prefill};
    \node[barlabel,anchor=south] at (2.785,.20) {Prefill};

    \draw[blocked] (.91,-.59) rectangle (2.31,-.27);
    \node[barlabel] at (1.61,-.43) {Blocked};
    \fill[submit] (.012,-.63) rectangle (.055,-.23);
    \fill[submit] (2.387,-.63) rectangle (2.43,-.23);

    \draw[\copystyle] (.84,-1.02) rectangle (2.31,-.70);
    \node[barlabel] at (1.575,-.86) {\copylabel};

    \draw[axis] (0,-1.18) -- (3.19,-1.18);
    \node[anchor=north east,inner sep=0pt,font=\figurefont\fontsize{6.8}{7.8}\selectfont]
      at (3.14,-1.22) {Time};
  \end{scope}
}
\end{tikzpicture}
\endgroup

%% file: tables/cuda-api-locks.tex
\begin{table}[t]
\centering
\caption{Observed C/S/T acquisitions by foreground CUDA API classes under CC.}
\label{tab:cuda-api-locks}
\small
\setlength{\tabcolsep}{4pt}
\renewcommand{\arraystretch}{1.15}
\begin{tabular*}{\columnwidth}{@{\extracolsep{\fill}}p{0.66\columnwidth}l@{}}
\toprule
\textbf{Foreground API class} & \textbf{Locks acquired} \\
\midrule
Event/stream creation and destruction, CUDA device/host allocation/free & $C_{\mathrm{W}}$ \\
Kernel/graph launches, event recording & $C_{\mathrm{R}} + S$ \\
Pinned H2D and D2H copies & $C_{\mathrm{R}} + S + T$ \\
\bottomrule
\end{tabular*}
\end{table}

%% file: sections/05-design-revised.tex
\section{\sysname Design}
\label{sec:design}

\sysname organizes KV transfers to reduce interference with
ready requests and accelerate restoration for requests awaiting
host-cached KV.
It builds on \emph{\baseline}
(\secref{sec:motivation:strawman}), retaining overlap scheduling
and deferring requests awaiting restored KV~\cite{vllmoffload}.

\input{figures/design-overview}

\figref{fig:design-overview} provides an overview of \sysname,
which combines three mechanisms (P1--P3) to coordinate
KV transfers with ongoing serving.
Across both transfer paths, \sysname manages resource lifetimes
to reduce scheduler-side stalls (P1).
For write-through backups, it bounds submissions and prioritizes
ready token readback (P2).
For restores that gate request admission, it uses auxiliary
contexts to reduce driver-lock interference and parallelize
copies (P3).

\subsection{Managing Resource Lifetimes (P1)}
\label{sec:design:resources}

Bulk copies hold C in read mode, so event lifecycle operations
and host allocations requiring write access can stall the scheduler.
\sysname reduces these conflicts by reusing resources across
batches and controlling their lifetimes.

For transfer synchronization, \sysname precreates readiness events.
The scheduler records an event after GPU computation prepares
a transfer's source data, and the copy worker waits for it
before copying.
Recording an existing event takes compatible read access to C,
while a separate stream avoids contention on the copy's S instance.
The event is reused after the corresponding worker wait completes,
avoiding creation and destruction on the scheduler path.

The same reuse policy applies to per-batch metadata, such as
sequence lengths.
The scheduler uploads this metadata from preallocated pinned
host buffers that become reusable after their uploads complete,
avoiding repeated host allocations.

Event operations can also arise indirectly when the allocator
reclaims temporary GPU arrays.
Gather and scatter use GPU index arrays to locate KV pages,
and \baseline applies allocator stream tracking to prevent
premature storage reuse.
The background workers already retain references to these arrays
until the operations using them complete.
\sysname uses these references to protect the arrays' lifetimes
and omits the redundant tracking, removing cleanup event
operations that can wait on C or S.

\subsection{Scheduling KV Backups (P2)}
\label{sec:design:wt}
\label{sec:characterization:granularity}

Token readback and KV backups acquire the same D2H transfer
lock T in the serving context, even on separate streams.
During result processing, the scheduler waits for token readback
to complete.
A long backup call can therefore stall subsequent batch scheduling
even after the GPU has produced the output tokens.
\sysname controls both the size of each backup submission
and the order in which the backup and response workers submit
their copies.

\bheading{Bounding backup submissions.}
To limit how much backup work is admitted at once while
retaining transfer bandwidth, we distinguish contiguous-buffer
size from the total bytes submitted per API call.
Grouping fewer whole buffers reduces submission size without
changing their individual sizes.
We vary these dimensions independently using
\texttt{cudaMemcpyBatchAsync} on H200 under CC with pinned
host buffers (\appref{sec:appendix:design}).

\input{figures/copy-granularity}

With 768~MiB per call, increasing buffer size from 128~KiB
to 6~MiB more than doubles bandwidth in both directions
(\figref{fig:copy-granularity}(a)).
Keeping buffers at 6~MiB while reducing bytes per call from
768 to 6~MiB preserves approximately 8~GB/s bandwidth,
but reduces median host-call duration from 98--99~ms
to 0.75--0.78~ms (\figref{fig:copy-granularity}(b)).
Smaller submissions therefore let the runtime reconsider
backup admission more frequently while retaining bandwidth.
\sysname gathers KV pages into GPU staging and copies whole
pages in groups, with a byte budget limiting pages per submission while preserving KV page size and layout.

\bheading{Prioritizing ready readback.}
Smaller submissions create more opportunities for readback,
but releasing T alone does not give the response worker
priority over the next backup submission.
In a probe with consecutive 6~MiB D2H submissions,
individual submissions complete in a median of 0.70~ms,
yet readback completion latency reaches a 90th percentile
of 106.4~ms.
Multiple backup groups complete while readback remains pending.

\sysname therefore coordinates submission order through a CPU
gate shared by the backup and response workers.
The response worker requests priority only after its readiness
event confirms that the GPU has produced the output tokens.
Until then, backups can continue.
Each worker holds the gate until its copy completes, limiting
backups to one group in flight.
When the gate is released, a waiting readback proceeds before
the next backup group.

\subsection{Parallel KV Restoration (P3)}
\label{sec:design:lb}
\label{sec:characterization:contexts}

Requests awaiting host-cached KV become eligible for admission
after restoration completes.
Bulk H2D copies in the serving context can also delay metadata
uploads needed by other ready batches.
\sysname uses auxiliary contexts to parallelize restores and
separate their C and T locks from foreground operations,
building on context parallelism for model-weight
loading~\cite{serializedbridge}.
For online restoration alongside serving and write-through
backups, \sysname coordinates host-allocation ownership and
transfer capacity.
Shared GPU staging across contexts lets the serving context
scatter restored KV directly, avoiding an intermediate assembly copy.

\bheading{Enabling concurrent copies.}
The shared host KV pool's allocation also affects copy parallelism.
On H200 under CC, 6~MiB H2D copies from a shared
\texttt{cudaHostAlloc} pool created in the serving context
remain serialized across copy contexts
(\figref{fig:context-parallelism}(a)).
With ordinary CPU memory, increasing the context count from
one to eight raises H2D bandwidth from 8.18 to 23.76~GB/s.
\sysname therefore allocates the host KV pool in ordinary CPU
memory, with each worker issuing explicit protected H2D copies
through a persistent auxiliary context and stream. Measurements with ordinary host memory also confirm that
bounded D2H submissions retain bandwidth and that readback
priority reduces foreground delays
(\appref{sec:appendix:design}).

\input{figures/context-parallelism}

\bheading{Allocating copy capacity.}
Concurrent backups still compete with restores for transfer
capacity.
With four H2D contexts, increasing D2H contexts from one to
eight improves D2H bandwidth but reduces H2D bandwidth by 23\%
(\figref{fig:context-parallelism}(b)).
Restores gate request admission, while write-through backups
can begin before GPU pages must be reclaimed.
\sysname therefore uses multiple restore contexts and a single backup worker in the serving context.
Keeping backups in this context also avoids consuming additional
GPU memory that could otherwise hold KV state.

\bheading{Shared staging and pipelined restoration.}
Restore workers copy host-cached KV into a persistent GPU staging
allocation, mapped into all copy contexts and the serving context
through CUDA virtual-memory interfaces.
Workers fill disjoint ranges, and scatter in the serving context
moves KV directly from these ranges into the GPU KV pool,
avoiding an additional assembly copy.

To overlap H2D copies with scatter, \sysname divides this allocation
into two regions (A/B in \figref{fig:design-overview}).
Workers fill one region while the serving context scatters the other.
Reusable completion events prevent a region from being
overwritten before its scatter finishes.

Within each region, scatter waits for all H2D copies to complete,
so uneven worker completion can stall the pipeline.
To balance work, \sysname divides contiguous host KV ranges
into bounded copy units in a shared queue.
Workers take the next available unit, allowing faster
workers to process more units as copy service times vary.
After the final scatter, dependent requests become
eligible for admission.

%% file: figures/design-overview.tex
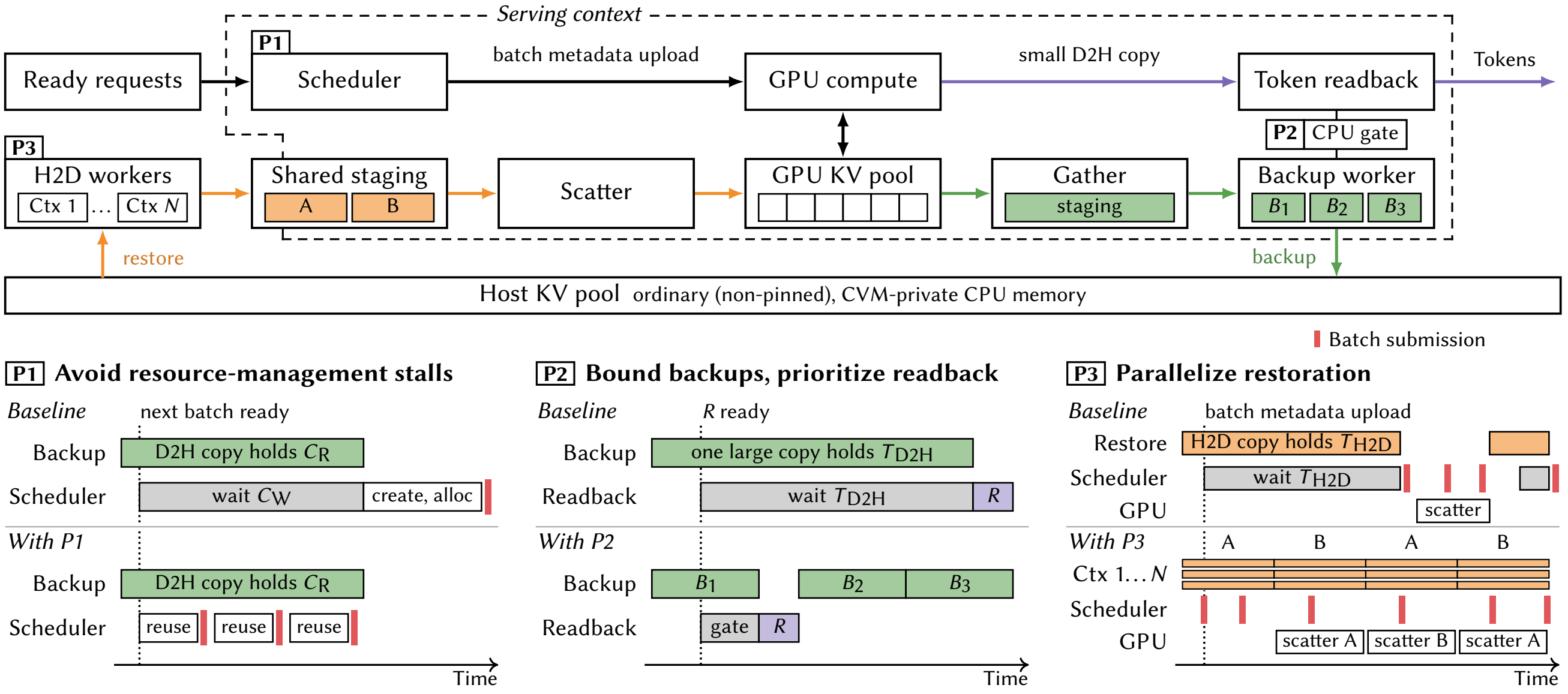
\begin{figure*}[t]
\centering
\newcommand{\figVsub}[1]{\raisebox{-.32ex}{\fontsize{6.5}{7}\selectfont #1}}
\newcommand{\figVlock}[2]{\textit{#1}\figVsub{#2}}
\newcommand{\figVsm}{\fontsize{6.8}{7.8}\selectfont}
\newcommand{\figVbar}[5]{%
  \draw[#1] (#2,#4-.16) rectangle (#3,#4+.16);
  \path (#2,#4) -- (#3,#4) node[midway,fvbarlabel] {#5};}
\newcommand{\figVop}[4]{%
  \draw[fvop] (#1,#3-.16) rectangle (#2,#3+.16);
  \path (#1,#3) -- (#2,#3) node[midway,fvoplabel] {#4};}
\newcommand{\figVwait}[4]{%
  \draw[fvwait] (#1,#3-.16) rectangle (#2,#3+.16);
  \ifx\relax#4\relax\else
    \path (#1,#3) -- (#2,#3) node[midway,fvbarlabel] {#4};\fi}
\newcommand{\figVtick}[2]{\fill[VeloCCHighlight] (#1-.035,#2-.2) rectangle (#1+.035,#2+.2);}
\newcommand{\figVdash}[4]{%
  \pgfmathsetmacro\figVlen{veclen((#3)-(#1),(#4)-(#2))*28.45274}%
  \pgfmathsetmacro\figVoff{(\figVlen-2.4)/max(floor((\figVlen-2.4)/5),1)-2.4}%
  \draw[line width=.6pt,line cap=rect,dash pattern=on 2.4pt off \figVoff pt] (#1,#2) -- (#3,#4);}
\newcommand{\figVpanel}[8]{%
  \ifnum#1=3\relax
    \def\figVbaselineOne{2.77}\def\figVbaselineTwo{2.37}%
    \def\figVwithOne{1.28}\def\figVwithTwo{.88}%
  \else
    \def\figVbaselineOne{2.66}\def\figVbaselineTwo{2.16}%
    \def\figVwithOne{1.17}\def\figVwithTwo{.67}%
  \fi
  \draw[draw=VeloCCInk!42,line width=.4pt] (-1.20,1.82) -- (4.235,1.82);
  \node[fvtab,anchor=base west] (t#1) at (-1.20,3.48) {P#1};
  \node[fvtitle] at ([xshift=.1cm]t#1.base east) {#2};
  \node[fvrow] at (-1.20,3.14) {Baseline};
  \node[fvrow] at (-1.20,1.65) {With P#1};
  \node[fvlane] at (-.08,\figVbaselineOne) {#3};
  \node[fvlane] at (-.08,\figVbaselineTwo) {#4};
  \node[fvlane] at (-.08,\figVwithOne) {#5};
  \node[fvlane] at (-.08,\figVwithTwo) {#6};
  \draw[fvguide] (#7,.25) -- (#7,1.77);
  \draw[fvguide] (#7,1.87) -- (#7,2.97);
  \node[fvsmall,inner sep=0pt,anchor=mid west] at (#7,3.14) {#8};
  \draw[fvaxis] (0,.25) -- (4.235,.25);
  \node[fvsmall,anchor=north east,inner sep=0pt] at (4.235,.19) {Time};}

\begin{tikzpicture}[
  x=1cm,y=0.97cm,name prefix=fv-,
  font=\figurefont\fontsize{8}{9.2}\selectfont,
  >={Latex[length=1.7mm,width=1.3mm]},
  fvbox/.style={draw=black,line width=.7pt,fill=white,minimum width=2.15cm,
    minimum height=.62cm,inner sep=1pt,align=center,text height=5.5pt,text depth=1.9pt},
  fvarr/.style={->,line width=.95pt,draw=black},
  fvR/.style={fvarr,draw=VeloCCCipher},
  fvB/.style={fvarr,draw=VeloCCBackup},
  fvT/.style={fvarr,draw=VeloCCToken},
  fvcell/.style={draw=black,line width=.5pt,minimum height=.31cm,inner sep=0pt,
    text height=4.7pt,text depth=1.65pt,font=\figurefont\fontsize{6.8}{7.4}\selectfont},
  fvtab/.style={draw=black,line width=.6pt,fill=white,inner xsep=2pt,inner ysep=1.2pt,
    font=\figurefont\bfseries\fontsize{7}{8}\selectfont},
  fvsmall/.style={font=\figurefont\fontsize{6.8}{7.8}\selectfont},
  fvlane/.style={anchor=mid east,inner sep=0pt,font=\figurefont\fontsize{7.5}{8.5}\selectfont},
  fvrow/.style={anchor=mid west,inner sep=0pt,font=\figurefont\itshape\fontsize{7.5}{8.5}\selectfont},
  fvtitle/.style={anchor=base west,inner sep=0pt,font=\figurefont\bfseries\fontsize{8}{9}\selectfont},
  fvbarlabel/.style={inner sep=0pt,text height=4.5pt,text depth=1.3pt,
    font=\figurefont\fontsize{6.8}{7.4}\selectfont},
  fvoplabel/.style={inner sep=0pt,text height=4.3pt,text depth=1.25pt,
    font=\figurefont\fontsize{6.5}{7.2}\selectfont},
  fvwait/.style={draw=black,line width=.5pt,fill=VeloCCInk!22},
  fvcopyR/.style={draw=black,line width=.5pt,fill=VeloCCCipher!60},
  fvcopyB/.style={draw=black,line width=.5pt,fill=VeloCCBackup!55},
  fvread/.style={draw=black,line width=.5pt,fill=VeloCCToken!45},
  fvop/.style={draw=black,line width=.5pt,fill=white},
  fvguide/.style={densely dotted,line width=.65pt},
  fvaxis/.style={-{To[length=1mm,width=1.6mm]},line width=.6pt},
  fvgate/.style={draw=black,line width=.6pt,fill=white,outer sep=0pt,minimum height=.32cm,
    inner ysep=0pt,text height=4.7pt,text depth=1.65pt},
]
\path[use as bounding box] (0,-.02) rectangle (17.145,7.77);

\begin{scope}[yshift=3mm]
\node[inner xsep=3pt,anchor=west,font=\figurefont\itshape\fontsize{7.5}{8.5}\selectfont] (ctxlabel) at (5.30,7.32) {Serving context};
\pgfextractx{\dimen0}{\pgfpointanchor{fv-ctxlabel}{east}}
\pgfmathsetmacro\figVlabx{\the\dimen0/1cm}
\figVdash{2.435}{5.97}{2.435}{6.49}
\figVdash{2.435}{6.65}{2.435}{7.32}
\figVdash{2.435}{7.32}{5.30}{7.32}
\figVdash{\figVlabx}{7.32}{15.95}{7.32}
\figVdash{15.95}{7.32}{15.95}{6.65}
\figVdash{15.95}{6.49}{15.95}{4.78}
\figVdash{15.95}{4.78}{14.755}{4.78}
\figVdash{14.595}{4.78}{3.06}{4.78}
\draw[line width=.6pt,line cap=rect] (3.06,4.78) -- (3.06,4.92);
\figVdash{3.06}{5.68}{3.06}{5.97}
\figVdash{3.06}{5.97}{2.435}{5.97}

\node[fvbox] (ready) at (1.075,6.57) {Ready requests};
\node[fvbox] (sched) at (3.795,6.57) {Scheduler};
\node[fvbox] (comp)  at (9.235,6.57) {GPU compute};
\node[fvbox] (read)  at (14.675,6.57) {Token readback};
\draw[fvarr] (ready) -- (sched);
\draw[fvarr] (sched) -- (comp) node[midway,above=1pt,fvsmall] {batch metadata upload};
\draw[fvT] (comp) -- (read) node[midway,above=1pt,fvsmall] {small D2H copy};
\draw[fvT] (read.east) -- (17.10,6.57) node[pos=.574,above=1pt,fvsmall] {Tokens};%

\node[fvbox,minimum height=.76cm] (work) at (1.075,5.30) {};
\node[anchor=base] at (1.075,5.42) {H2D workers};
\node[fvcell,minimum width=.76cm] at (0.525,5.14) {Ctx 1};
\node[fvcell,draw=none] at (1.075,5.14) {\dots};
\node[fvcell,minimum width=.76cm] at (1.625,5.14) {Ctx \textit{N}};

\node[fvbox,minimum height=.76cm] (stag) at (3.795,5.30) {};
\node[anchor=base] at (3.795,5.42) {Shared staging};
\node[fvcell,minimum width=.90cm,fill=VeloCCCipher!60] at (3.315,5.14) {A};
\node[fvcell,minimum width=.90cm,fill=VeloCCCipher!60] at (4.275,5.14) {B};

\node[fvbox,minimum height=.76cm] (scat) at (6.515,5.30) {Scatter};

\node[fvbox,minimum height=.76cm] (pool) at (9.235,5.30) {};
\node[anchor=base] at (9.235,5.42) {GPU KV pool};
\foreach \i in {0,...,5} {\draw[line width=.5pt] (8.305+.31*\i,4.985) rectangle ++(.31,.31);}

\node[fvbox,minimum height=.76cm] (gath) at (11.955,5.30) {};
\node[anchor=base] at (11.955,5.42) {Gather};
\node[fvcell,minimum width=1.86cm,fill=VeloCCBackup!55] at (11.955,5.14) {staging};

\node[fvbox,minimum height=.76cm] (bkw) at (14.675,5.30) {};
\node[anchor=base] at (14.675,5.42) {Backup worker};
\node[fvcell,minimum width=.58cm,fill=VeloCCBackup!55] at (14.035,5.14) {\textit{B}\figVsub{1}};
\node[fvcell,minimum width=.58cm,fill=VeloCCBackup!55] at (14.675,5.14) {\textit{B}\figVsub{2}};
\node[fvcell,minimum width=.58cm,fill=VeloCCBackup!55] at (15.315,5.14) {\textit{B}\figVsub{3}};

\draw[fvR] (work) -- (stag);
\draw[fvR] (stag) -- (scat);
\draw[fvR] (scat) -- (pool);
\draw[fvB] (pool) -- (gath);
\draw[fvB] (gath) -- (bkw);
\draw[fvarr,<->] (comp.south) -- (pool.north);

\node[fvgate,inner xsep=2.5pt,fvsmall] (gate) at (14.882,5.97) {CPU gate};
\draw[line width=.6pt] (read.south) -- (read.south |- gate.north);
\draw[line width=.6pt] (bkw.north) -- (bkw.north |- gate.south);

\node[fvbox,outer sep=0pt,minimum width=17.145cm,minimum height=.40cm,anchor=west] (host) at (0,4.14)
  {Host KV pool\enspace{\figVsm ordinary (non-pinned), CVM-private CPU memory}};
\draw[fvR] (1.075,4.34) -- (work.south);
\draw[fvB] (bkw.south) -- (14.675,4.34);
\node[fvsmall,anchor=west,text=VeloCCCipher!85!black] at (1.17,4.56) {restore};
\node[fvsmall,anchor=east,text=VeloCCBackup!80!black] at (14.58,4.56) {backup};

\node[fvtab,anchor=south west,outer sep=0pt] at ([xshift=.35pt,yshift=-.35pt]sched.north west) {P1};
\node[fvtab,fvgate,anchor=east] at (gate.west) {P2};
\node[fvtab,anchor=south west,outer sep=0pt] at ([xshift=.35pt,yshift=-.35pt]work.north west) {P3};
\end{scope}

\fill[VeloCCHighlight] (14.43,3.86) rectangle (14.49,4.05);
\node[fvsmall,anchor=mid west,inner sep=0pt] at (14.58,3.955) {Batch submission};

\begin{scope}[shift={(1.20,0)}]
  \figVpanel{1}{Avoid resource-management stalls}{Backup}{Scheduler}{Backup}{Scheduler}{.28}{next batch ready}
  \figVbar{fvcopyB}{.08}{2.75}{2.66}{D2H copy holds \figVlock{C}{R}}
  \figVwait{.28}{2.75}{2.16}{wait \figVlock{C}{W}}
  \figVop{2.75}{4.05}{2.16}{create, alloc}
  \figVtick{4.125}{2.16}
  \figVbar{fvcopyB}{.08}{2.75}{1.17}{D2H copy holds \figVlock{C}{R}}
  \figVop{.28}{.92}{.67}{reuse}
  \figVtick{.99}{.67}
  \figVop{1.11}{1.75}{.67}{reuse}
  \figVtick{1.82}{.67}
  \figVop{1.94}{2.58}{.67}{reuse}
  \figVtick{2.65}{.67}
\end{scope}

\begin{scope}[shift={(7.048,0)}]
  \figVpanel{2}{Bound backups, prioritize readback}{Backup}{Readback}{Backup}{Readback}{.62}{\textit{R} ready}
  \figVbar{fvcopyB}{.08}{3.62}{2.66}{one large copy holds \figVlock{T}{D2H}}
  \figVwait{.62}{3.62}{2.16}{wait \figVlock{T}{D2H}}
  \figVbar{fvread}{3.62}{4.06}{2.16}{\textit{R}}
  \figVbar{fvcopyB}{.08}{1.26}{1.17}{\textit{B}\figVsub{1}}
  \figVbar{fvcopyB}{1.70}{2.88}{1.17}{\textit{B}\figVsub{2}}
  \figVbar{fvcopyB}{2.88}{4.06}{1.17}{\textit{B}\figVsub{3}}
  \figVwait{.62}{1.26}{.67}{gate}
  \figVbar{fvread}{1.26}{1.70}{.67}{\textit{R}}
\end{scope}

\begin{scope}[shift={(12.897,0)}]
  \renewcommand{\figVbar}[5]{%
    \draw[#1] (#2,#4-.13) rectangle (#3,#4+.13);
    \path (#2,#4) -- (#3,#4) node[midway,fvbarlabel] {#5};}
  \renewcommand{\figVop}[4]{%
    \draw[fvop] (#1,#3-.13) rectangle (#2,#3+.13);
    \path (#1,#3) -- (#2,#3) node[midway,fvoplabel] {#4};}
  \renewcommand{\figVwait}[4]{%
    \draw[fvwait] (#1,#3-.13) rectangle (#2,#3+.13);
    \ifx\relax#4\relax\else
      \path (#1,#3) -- (#2,#3) node[midway,fvbarlabel] {#4};\fi}
  \renewcommand{\figVtick}[2]{%
    \fill[VeloCCHighlight] (#1-.035,#2-.16) rectangle (#1+.035,#2+.16);}

  \figVpanel{3}{Parallelize restoration}{Restore}{Scheduler}{Ctx 1\dots\textit{N}}{Scheduler}{.32}{batch metadata upload}
  \node[fvlane] at (-.08,2.00) {GPU};
  \node[fvlane] at (-.08,.51) {GPU};

  \figVbar{fvcopyR}{.08}{2.48}{2.77}{H2D copy holds \figVlock{T}{H2D}}
  \figVbar{fvcopyR}{3.465}{4.12}{2.77}{}
  \figVwait{.32}{2.48}{2.37}{wait \figVlock{T}{H2D}}
  \figVtick{2.555}{2.37}
  \figVtick{3.00}{2.37}
  \figVtick{3.39}{2.37}
  \figVwait{3.80}{4.12}{2.37}{}
  \figVtick{4.195}{2.37}
  \figVop{2.665}{3.465}{2.00}{scatter}

  \fill[white] (.08,1.28-.18) rectangle (4.12,1.28+.18);
  \foreach \dy in {-.125,0,.125}
    \foreach \xa/\xb in {.08/1.09,1.09/2.10,2.10/3.11,3.11/4.12}
      \draw[fvcopyR,line width=.4pt]
        (\xa,1.28+\dy-.0425) rectangle (\xb,1.28+\dy+.0425);
  \foreach \x/\r in {.585/A,1.595/B,2.605/A,3.615/B}
    \node[fvbarlabel,anchor=mid] at (\x,1.65) {\r};

  \foreach \t in {.32,.74,1.50,2.50,3.50,4.10} {
    \figVtick{\t}{.88}}
  \figVop{1.115}{2.075}{.51}{scatter A}
  \figVop{2.125}{3.085}{.51}{scatter B}
  \figVop{3.135}{4.095}{.51}{scatter A}
\end{scope}
\end{tikzpicture}
\caption{\sysname overview and baseline/P1--P3 timelines. Gray: waits.
$C_{\mathrm{R}}$/$C_{\mathrm{W}}$: context-lock read/write access.
$T$: per-context, per-direction transfer lock. A/B: staging regions.
$B_i$: backup groups. $R$: token readback.}
\label{fig:design-overview}
\end{figure*}

%% file: figures/copy-granularity.tex
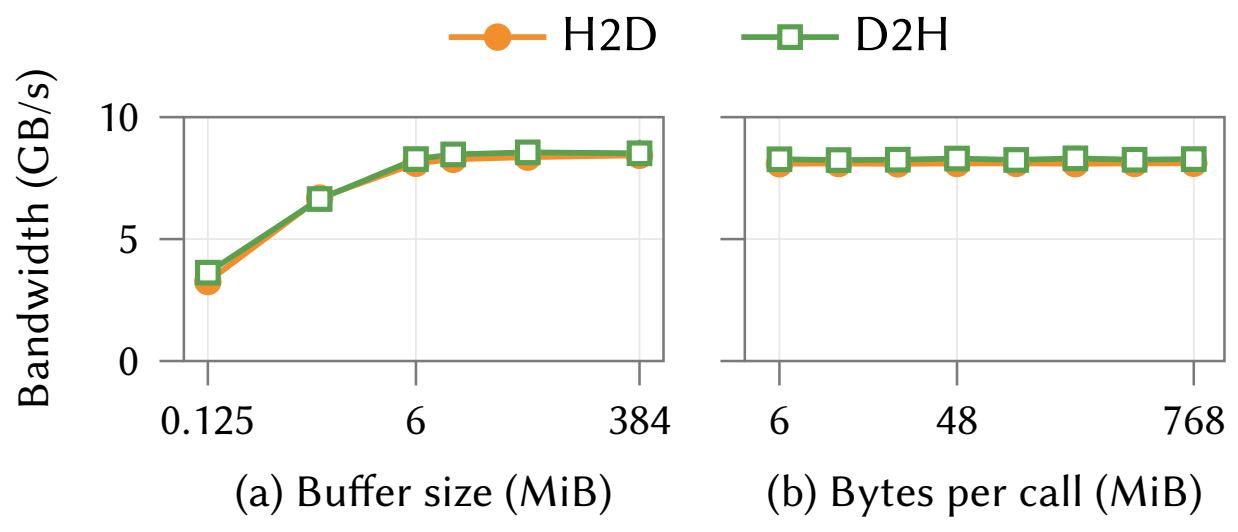
\begin{figure}[t]
\centering
\input{figures/copy-granularity-plots}
\caption{Transfer bandwidth with the batched API and pinned host buffers on H200 under CC. (a) Vary contiguous-buffer size with 768~MiB per call. (b) Vary bytes per call with 6~MiB buffers.}
\label{fig:copy-granularity}
\end{figure}

%% file: figures/copy-granularity-plots.tex
\resizebox{\columnwidth}{!}{%
\begin{tikzpicture}[font=\figurefont\fontsize{9}{10}\selectfont]
\pgfplotsset{
    granularity axis/.style={
        /pgfplots/.cd,
        scale only axis, width=2.95cm, height=1.50cm,
        axis line style={VeloCCInk!65, line width=.45pt},
        tick style={VeloCCInk!65, line width=.45pt}, tick align=outside, tick pos=left,
        tick label style={font=\figurefont\fontsize{8}{9}\selectfont},
        label style={font=\figurefont\fontsize{8}{9}\selectfont},
        log basis x=2, minor x tick num=0, scaled ticks=false,
        grid=major, grid style={VeloCCInk!12, line width=.3pt},
        ymin=0, ymax=10, ytick={0,5,10}, yticklabels={0,5,10},
        clip=true,
    },
    granularity h2d/.style={VeloCCCipher, line width=1pt, mark=*, mark size=1.9pt},
    granularity d2h/.style={VeloCCBackup, line width=1pt, mark=square*, mark size=1.8pt,
                           mark options={fill=white, line width=.9pt}},
    granularity errors/.style={error bars/.cd, y dir=both, y explicit,
                               error bar style={line width=.55pt},
                               error mark options={mark size=1.6pt, line width=.55pt},
                               /pgfplots/.cd},
}
\begin{groupplot}[
    granularity axis,
    group style={group size=2 by 1, horizontal sep=.50cm},
]
    \nextgroupplot[
        xmode=log, xlabel={(a) Buffer size (MiB)}, ylabel={Bandwidth (GB/s)},
        xmin=.08, xmax=600,
        xtick={.125,6,384}, xticklabels={0.125,6,384},
        legend to name=copyGranularityLegend, legend columns=2,
        legend style={draw=none, font=\figurefont\fontsize{9}{10}\selectfont,
                      /tikz/every even column/.append style={column sep=12pt},
                      cells={anchor=west}},
    ]
    \addplot[granularity h2d,granularity errors]
        table[x=buffer_mib,y=batch_gbps,y error minus=batch_gbps_minus,
              y error plus=batch_gbps_plus,col sep=comma,row sep=\\] {
buffer_mib,runs,samples_per_run,batch_gbps,batch_gbps_minus,batch_gbps_plus,scalar_gbps,scalar_gbps_minus,scalar_gbps_plus\\
0.125000,3,5,3.260400,0.012500,0.002400,3.263400,0.015500,0.000600\\
1.000000,3,5,6.682000,0.009800,0.006700,6.696900,0.033200,0.002100\\
6.000000,3,5,8.107500,0.016600,0.170900,8.114200,0.029300,0.157700\\
12.000000,3,5,8.271300,0.022100,0.172500,8.270000,0.020900,0.107800\\
48.000000,3,5,8.358900,0.002200,0.252800,8.370600,0.055800,0.246100\\
384.000000,3,5,8.434600,0.048500,0.102700,8.418800,0.021500,0.235900\\
};
    \addlegendentry{H2D}
    \addplot[granularity d2h,granularity errors]
        table[x=buffer_mib,y=batch_gbps,y error minus=batch_gbps_minus,
              y error plus=batch_gbps_plus,col sep=comma,row sep=\\] {
buffer_mib,runs,samples_per_run,batch_gbps,batch_gbps_minus,batch_gbps_plus,scalar_gbps,scalar_gbps_minus,scalar_gbps_plus\\
0.125000,3,5,3.622500,0.015900,0.009400,3.570600,0.017600,0.015600\\
1.000000,3,5,6.642000,0.001200,0.017700,6.643900,0.013600,0.006300\\
6.000000,3,5,8.280000,0.063600,0.014700,8.260100,0.081800,0.003800\\
12.000000,3,5,8.472600,0.094300,0.052200,8.395800,0.034500,0.132000\\
48.000000,3,5,8.556200,0.097000,0.016500,8.511200,0.111700,0.068600\\
384.000000,3,5,8.518500,0.088800,0.179000,8.764400,0.043600,0.089100\\
};
    \addlegendentry{D2H}
    \nextgroupplot[
        xmode=log, xlabel={(b) Bytes per call (MiB)},
        xmin=4, xmax=1100, xtick={6,48,768}, xticklabels={6,48,768},
        yticklabels=\empty,
    ]
    \addplot[granularity h2d,granularity errors]
        table[x=call_mib,y=batch_gbps,y error minus=batch_gbps_minus,
              y error plus=batch_gbps_plus,col sep=comma,row sep=\\] {
call_mib,runs,samples_per_run,batch_gbps,batch_gbps_minus,batch_gbps_plus,batch_call_ms,batch_call_ms_minus,batch_call_ms_plus,sync_gbps,sync_gbps_minus,sync_gbps_plus,sync_call_ms,sync_call_ms_minus,sync_call_ms_plus\\
6,3,5,8.089100,0.002700,0.095000,0.779462,0.011459,0.000710,8.054900,0.009600,0.185200,0.733411,0.020407,0.002152\\
12,3,5,8.098700,0.018700,0.178300,1.553474,0.036289,0.000711,8.095400,0.036900,0.162200,1.505664,0.032329,0.004562\\
24,3,5,8.084500,0.057000,0.191800,3.107789,0.074930,0.018400,8.085400,0.011600,0.180500,3.060340,0.068720,0.002919\\
48,3,5,8.100800,0.018900,0.174400,6.202488,0.128398,0.016040,8.106300,0.029900,0.071100,6.153199,0.054190,0.019059\\
96,3,5,8.103300,0.024600,0.084000,12.425315,0.135289,0.026001,8.119700,0.045200,0.159300,12.333385,0.242836,0.064830\\
192,3,5,8.090400,0.006000,0.206400,24.865421,0.600524,0.015120,8.099700,0.025100,0.177300,24.832560,0.545645,0.027671\\
384,3,5,8.101800,0.020700,0.109600,49.702210,0.642684,0.126868,8.112000,0.040400,0.163700,49.592922,0.955180,0.268928\\
768,3,5,8.107500,0.016600,0.170900,99.337601,1.986408,0.219379,8.100300,0.016800,0.093500,99.437212,1.178417,0.214968\\
};
    \addplot[granularity d2h,granularity errors]
        table[x=call_mib,y=batch_gbps,y error minus=batch_gbps_minus,
              y error plus=batch_gbps_plus,col sep=comma,row sep=\\] {
call_mib,runs,samples_per_run,batch_gbps,batch_gbps_minus,batch_gbps_plus,batch_call_ms,batch_call_ms_minus,batch_call_ms_plus,sync_gbps,sync_gbps_minus,sync_gbps_plus,sync_call_ms,sync_call_ms_minus,sync_call_ms_plus\\
6,3,5,8.273600,0.058500,0.019500,0.749112,0.000120,0.006760,8.180100,0.061900,0.062500,0.753211,0.001389,0.008270\\
12,3,5,8.240300,0.025000,0.014000,1.523264,0.000938,0.002081,8.239600,0.022200,0.038500,1.517284,0.006399,0.013190\\
24,3,5,8.255800,0.034400,0.033200,3.039340,0.000761,0.001321,8.289900,0.025100,0.062700,3.017570,0.024390,0.017749\\
48,3,5,8.303100,0.014600,0.002300,6.090288,0.022179,0.029251,8.270300,0.049800,0.029700,6.071989,0.073079,0.016541\\
96,3,5,8.245000,0.012400,0.088300,12.140678,0.050290,0.030220,8.270500,0.010700,0.065900,12.107969,0.051250,0.067259\\
192,3,5,8.311400,0.086800,0.027000,24.329945,0.208618,0.295389,8.273500,0.022100,0.021200,24.407855,0.132890,0.153108\\
384,3,5,8.254200,0.010000,0.007300,48.888649,0.176298,0.201988,8.312400,0.015300,0.041700,48.453953,0.143288,0.366917\\
768,3,5,8.280000,0.063600,0.014700,97.692399,0.080779,1.245498,8.286800,0.035300,0.027700,97.479331,0.142539,0.252817\\
};
\end{groupplot}
\path (group c1r1.north west) -- coordinate (legendcenter) (group c2r1.north east);
\node[anchor=south] at ([yshift=1mm]legendcenter)
    {\pgfplotslegendfromname{copyGranularityLegend}};
\end{tikzpicture}%
}

%% file: figures/context-parallelism.tex
\begin{figure}[t]
\centering
\input{figures/context-parallelism-plot}
\caption{Context parallelism on H200 under CC.
(a) Each direction alone.
(b) Four H2D contexts with varying D2H concurrency.}
\label{fig:context-parallelism}
\end{figure}
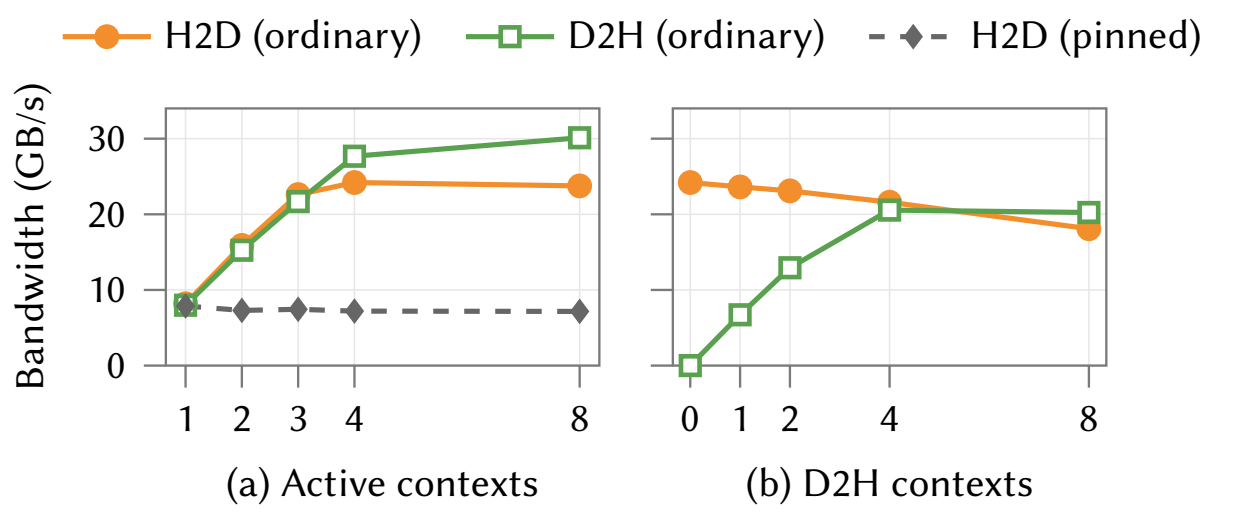

%% file: figures/context-parallelism-plot.tex
\resizebox{\columnwidth}{!}{%
\begin{tikzpicture}[font=\figurefont\fontsize{9}{10}\selectfont]
\pgfplotsset{
    context axis/.style={
        /pgfplots/.cd,
        scale only axis, width=2.95cm, height=1.75cm,
        axis line style={VeloCCInk!65, line width=.45pt},
        tick style={VeloCCInk!65, line width=.45pt}, tick align=outside, tick pos=left,
        tick label style={font=\figurefont\fontsize{8}{9}\selectfont},
        label style={font=\figurefont\fontsize{8}{9}\selectfont},
        scaled ticks=false, grid=major, grid style={VeloCCInk!12, line width=.3pt},
        ymin=0, ymax=34, ytick={0,10,20,30}, yticklabels={0,10,20,30},
        clip=true,
    },
    context h2d/.style={VeloCCCipher, line width=1pt, mark=*, mark size=1.9pt},
    context d2h/.style={VeloCCBackup, line width=1pt, mark=square*, mark size=1.8pt,
                       mark options={fill=white, line width=.9pt}},
    context pinned/.style={VeloCCInk!75, dashed, line width=1pt,
                          mark=diamond*, mark size=2pt,
                          mark options={solid, fill=VeloCCInk!75, line width=.6pt}},
}
\begin{groupplot}[
    context axis,
    group style={group size=2 by 1, horizontal sep=.50cm},
]
    \nextgroupplot[
        xlabel={(a) Active contexts}, ylabel={Bandwidth (GB/s)},
        xmin=.65, xmax=8.35, xtick={1,2,3,4,8}, xticklabels={1,2,3,4,8},
        legend to name=contextParallelismLegend, legend columns=3,
        legend style={draw=none, font=\figurefont\fontsize{8}{9}\selectfont,
                      /tikz/every even column/.append style={column sep=6pt},
                      cells={anchor=west}},
    ]
    \addplot[context h2d]
        table[x=active_contexts,y=h2d_gbps,col sep=comma,row sep=\\] {
active_contexts,h2d_gbps,d2h_gbps,pinned_h2d_gbps\\
1,8.177215,7.979244,7.875225\\
2,15.888024,15.247134,7.294734\\
3,22.612332,21.682874,7.439018\\
4,24.199456,27.679051,7.200781\\
8,23.756538,30.118458,7.160516\\
};
    \addlegendentry{H2D (ordinary)}
    \addplot[context d2h]
        table[x=active_contexts,y=d2h_gbps,col sep=comma,row sep=\\] {
active_contexts,h2d_gbps,d2h_gbps,pinned_h2d_gbps\\
1,8.177215,7.979244,7.875225\\
2,15.888024,15.247134,7.294734\\
3,22.612332,21.682874,7.439018\\
4,24.199456,27.679051,7.200781\\
8,23.756538,30.118458,7.160516\\
};
    \addlegendentry{D2H (ordinary)}
    \addplot[context pinned]
        table[x=active_contexts,y=pinned_h2d_gbps,col sep=comma,row sep=\\] {
active_contexts,h2d_gbps,d2h_gbps,pinned_h2d_gbps\\
1,8.177215,7.979244,7.875225\\
2,15.888024,15.247134,7.294734\\
3,22.612332,21.682874,7.439018\\
4,24.199456,27.679051,7.200781\\
8,23.756538,30.118458,7.160516\\
};
    \addlegendentry{H2D (pinned)}
    \nextgroupplot[
        xlabel={(b) D2H contexts},
        xmin=-.35, xmax=8.35, xtick={0,1,2,4,8}, xticklabels={0,1,2,4,8},
        yticklabels=\empty,
    ]
    \addplot[context h2d]
        table[x=d2h_contexts,y=h2d_gbps,col sep=comma,row sep=\\] {
d2h_contexts,h2d_gbps,d2h_gbps\\
0,24.199456,0.000000\\
1,23.622320,6.714242\\
2,23.098871,12.931878\\
4,21.595632,20.555445\\
8,18.092550,20.240033\\
};
    \addplot[context d2h]
        table[x=d2h_contexts,y=d2h_gbps,col sep=comma,row sep=\\] {
d2h_contexts,h2d_gbps,d2h_gbps\\
0,24.199456,0.000000\\
1,23.622320,6.714242\\
2,23.098871,12.931878\\
4,21.595632,20.555445\\
8,18.092550,20.240033\\
};
\end{groupplot}
\path (group c1r1.north west) -- coordinate (legendcenter) (group c2r1.north east);
\node[anchor=south] at ([yshift=1mm]legendcenter)
    {\pgfplotslegendfromname{contextParallelismLegend}};
\end{tikzpicture}%
}

%% file: sections/06-evaluation.tex
\section{Evaluation}
\label{sec:evaluation}

In this section, we quantify \sysname's end-to-end throughput and TTFT gains under CC (\secref{sec:evaluation:macro}) and explain its throughput gains and remaining gaps across workloads and models (\secref{sec:evaluation:gap}).
We then isolate each optimization's contribution (\secref{sec:evaluation:ablation}) and evaluate generality across CC GPU platforms (\secref{sec:evaluation:generalization}).

\subsection{Evaluation Setup}
\label{sec:evaluation:setup}
\label{sec:design:impl}

\bheading{Implementation and testbed.}
We implement \sysname in SGLang~0.5.18 with HiCache,
adding or modifying approximately 3,800 lines of runtime
code, excluding tests and evaluation scripts.
Our H200 experiments use a single NVIDIA H200 NVL GPU
connected through PCIe~5.0 in an AMD SEV-SNP confidential
VM with 16 vCPUs.
CC-on and CC-off runs share the same software stack:
NVIDIA driver~610.43.02, CUDA~13.0, and PyTorch~2.13.0.
We use a static GPU-memory fraction of 0.9 for model weights and the KV pool, with a 500~GB host cache when enabled.

\bheading{Models.}
We evaluate three models with 14B--106B total parameters:
Qwen2.5-14B-Instruct~\cite{qwen25},
Qwen3.8-27B~\cite{qwen38-27b},
and GLM-4.5-Air~\cite{glm45air}.
They cover dense full-attention, hybrid-attention, and
mixture-of-experts architectures, respectively.
The Qwen models use BF16 weights, while GLM uses a
4-bit AWQ checkpoint~\cite{glm45air-awq}.
All models use BF16 KV caches and a 131,072-token
context window.

\bheading{Serving configurations.}
\emph{GPU-only} retains GPU prefix caching but disables
the host cache.
\emph{\baseline} uses the staged KV-transfer workers
and deferred restore admission of
\secref{sec:motivation:strawman}.
\sysname adds P1--P3 to \baseline.
CC-off references are GPU-only and native HiCache
with kernel transfers and deferred restore admission.
All CC-on configurations enable the background response
worker, and all host-cache configurations use write-through
backups.
\sysname uses three auxiliary restore contexts and
32~MiB copy-group budgets.

\bheading{Workloads.}
Following prior work~\cite{strata,mori}, we evaluate
document-prefix sharing and multi-turn session reuse.
\emph{LooGLE}~\cite{loogle} and
\emph{NarrativeQA}~\cite{narrativeqa} issue independent
questions over shared documents, with Poisson arrivals
and at most 128 requests in flight.
For \emph{SWE-bench}, we replay LMCache agentic
traces~\cite{lmcache-agentic-traces} with a fixed pool
of active sessions.
Each follow-up turn arrives after the previous response
and the recorded inter-turn delay.
Completed sessions are replaced to maintain the pool.
\tabref{tab:evaluation-workloads} summarizes the prepared
workloads shared across all three models.

After compilation warm-up, LooGLE and SWE-bench start
with empty caches.
NarrativeQA starts with an empty GPU cache and, when
enabled, a host cache prepopulated with document KV.
This emphasizes restoration, complementing LooGLE's
initial prefill and cache population.
Prepopulation is excluded from measurement.

\input{tables/evaluation-workloads}

\bheading{Metrics.}
We report output-token throughput and mean time to first
token (TTFT) from client dispatch.
Document workloads replay the full request set once per
load, including the final drain.
SWE-bench runs for 1,200 seconds.
Throughput counts output tokens received during seconds
600--1,200, and TTFT covers requests arriving in that window.
Inter-token latency (ITL) results appear in
\appref{sec:appendix:itl}.

\subsection{Q1: How Much Does \sysname Improve Serving Performance under CC?}
\label{sec:evaluation:macro}

\input{figures/e2e-h200}

Across the nine model--workload pairs, \sysname raises throughput by a geometric mean of $1.74\times$ over \baseline and lowers mean TTFT by 36\% on average at each pair's \emph{peak load}, the highest load measured for all three host-cache configurations. \figref{fig:e2e-h200} shows throughput and mean TTFT as request rate or active-session count increases.

\bheading{Document workloads.}
Gains are largest on NarrativeQA, where the prepopulated host cache makes restoration prominent in the measured work: up to $2.7\times$ \baseline's throughput at peak load, with mean TTFT cut by more than half on all three models. LooGLE, where initial prefill and cache population dominate the measured work, gains less but improves on every model. GPU-only, with no host cache to reuse evicted KV, falls far behind at lower request rates on both workloads.

\bheading{Agent sessions.}
On SWE-bench, gains grow with the session pool. At small pools nearly all prompt tokens hit in the GPU cache and the three configurations perform alike. As the pool grows, GPU-only drops sharply, \baseline declines, and \sysname sustains higher throughput, reaching $1.15$--$1.36\times$ \baseline's throughput at 64 sessions with lower mean TTFT.

\bheading{Comparison with CC-off.}
At peak load, \sysname comes within 10\% of CC-off HiCache's throughput on LooGLE and SWE-bench. NarrativeQA is the exception, at 65--73\% of CC-off: it has both the largest gain over \baseline and the largest remaining gap. \secref{sec:evaluation:gap} explains both.

\subsection{Understanding Gains and Remaining Gaps}
\label{sec:evaluation:gap}

We combine measured serving behavior with a simple capacity
model to explain where \sysname helps most and what continues
to limit performance in \emph{restore-saturated} and
\emph{restore-unsaturated} regimes.

\bheading{Analysis framework.}
The model estimates \sysname's throughput relative to CC-off,
separating initial cache population from subsequent reuse.
For each phase $p$, let $c_p$ and $\rho_p$ denote estimated
compute and restoration times normalized by that phase's
CC-off duration.
The compute term includes additional prefill work under CC,
while the restore term includes copy time and waits between restores.
A demand below one fits within the reference duration.
Since computation and restoration can overlap, we approximate
each phase's slowdown by its larger resource demand.
Weighting by the phase's share $f_p$ of CC-off time gives
\begin{equation}
\widehat R =
\left[\sum_p f_p\max\{1,\,c_p,\,\rho_p\}\right]^{-1}.
\label{eq:cc-gap-model}
\end{equation}
A phase is \emph{restore-saturated} when
$\rho_p > \max\{1,c_p\}$, so restoration limits throughput.
Otherwise, it is \emph{restore-unsaturated}.
Across 63 paired configurations, the mean absolute error
is 1.51 percentage points.
Calibration and supporting measurements appear in
\appref{sec:appendix:gap-model}.

\input{figures/gap-mechanisms-h200}

\bheading{Restore-saturated regime.}
NarrativeQA's prepopulated host cache makes restoration
prominent in the measured work, giving faster restores
substantial scope to improve over \baseline.
Although \sysname improves restore bandwidth, the required
KV volume remains large relative to the effective
protected-transfer bandwidth.
At high load, the resulting copy service time and admission
waits together exceed the CC-off replay duration
(\figref{fig:gap-mechanisms-h200}(a)).
Restoration therefore continues to limit throughput,
explaining why large gains over \baseline coexist
with a large remaining gap to CC-off.

\bheading{Restore-unsaturated regime.}
SWE-bench generates about 212 tokens per turn, versus 6--15 tokens per request on document workloads. These longer generation phases spread restoration work over more serving time and allow it to overlap ongoing computation. Estimated restore demand is below 0.41 across measured SWE-bench loads, helping explain why \sysname approaches CC-off throughput.

With restoration below capacity, the model attributes
the remaining gap to compute overhead
(\figref{fig:gap-mechanisms-h200}(b)).
For the hybrid 27B model, forward-only probes show that
short prefill executions launch many small kernels
individually, making them sensitive to CC's higher
per-launch cost.
At 64 SWE-bench sessions, its estimated extra prefill work is 5.7\% of CC-off time, contributing to
the remaining gap even when restoration demand stays
below capacity.

\bheading{Effect of phase composition.}
LooGLE includes initial prefill and cache population,
so its gains depend on protecting foreground execution
during backups as well as accelerating restores.
At peak load, the initial phase accounts
for 74--83\% of CC-off time, with estimated extra prefill
work adding only 1--4\% of that phase's duration.
Restoration pressure during later reuse therefore affects
a smaller part of the replay, limiting the overall gap.

\subsection{Q2: Which Optimizations Drive the Gains?}
\label{sec:evaluation:ablation}

\input{tables/ablation-h200}

We evaluate cumulative P1--P3 additions and individual P3
components for Qwen2.5-14B on LooGLE at 12~req/s,
where \sysname's throughput has saturated (\tabref{tab:ablation-h200}).

\bheading{Protecting foreground execution (P1 and P2).}
P1 cuts resource-management stalls, raising throughput by 31\% over \baseline. P2 bounds backup submissions and prioritizes ready token readback, adding another 2\%.

\bheading{Accelerating restoration (P3).}
P3 base combines three auxiliary restore contexts with the
host-memory and shared-staging organization described in
\secref{sec:design:lb}, increasing throughput by 53\% over P1+P2.
Work balancing and H2D--scatter overlap together provide a further
6.4\% improvement.
The mechanism is bandwidth: restored bytes divided by summed restore durations rise from 6.8~GB/s without P3 to 20.0~GB/s with full P3.

\subsection{Generality Across CC GPU Platforms}
\label{sec:evaluation:generalization}

\input{tables/e2e-pro6000}

\bheading{Cross-platform setup.}
We repeat the evaluation on a single 96-GB NVIDIA RTX PRO 6000
Blackwell Server Edition GPU.
Both host CPUs support AVX-512; the RTX PRO 6000 host's Zen~5 cores
provide a 512-bit vector datapath, while the H200 host's Zen~4 cores
execute 512-bit operations in two 256-bit steps~\cite{amdzenfour,amdzenfive}.
We follow the setup in \secref{sec:evaluation:setup}, using 30
NarrativeQA documents to fit the host cache. Platform-specific
settings appear in \appref{sec:appendix:evaluation}.

\bheading{Serving performance.}
Across the nine model--workload pairs in \tabref{tab:e2e-pro6000}, \sysname achieves a geometric-mean throughput speedup of $1.28\times$ over \baseline and reduces mean TTFT by 19\% on average. Gains are smaller than on H200 because the protected-transfer path is faster: Zen~5's wider vector datapath may contribute to faster CPU-side encryption and the roughly doubled protected H2D bandwidth from about 8~GB/s on H200 to 17~GB/s, which lowers \baseline's restore cost and leaves less for \sysname to recover.

The driver-lock scopes match those of \secref{sec:characterization:locks}, and both regimes of \secref{sec:evaluation:gap} carry over. NarrativeQA remains \emph{restore-saturated}, with the largest speedups over \baseline, up to $2.2\times$, and the largest remaining gap to CC-off. SWE-bench remains \emph{restore-unsaturated}, within 5\% of CC-off throughput. The one pair without a gain, 27B on SWE-bench, is within 1\% of \baseline's throughput: the platform's single-context bandwidth already meets its restore demand, so auxiliary contexts add little throughput and cost 6.3\% of GPU KV capacity.

%% file: tables/evaluation-workloads.tex
\begin{table}[t]
\centering
\caption{Workloads shared across all three models. Token counts show ranges of per-request means across models. SWE-bench size refers to the full trace.}
\label{tab:evaluation-workloads}
\small
\setlength{\tabcolsep}{3pt}
\renewcommand{\arraystretch}{1.15}
\newcommand{\workloadcell}[2][l]{{\renewcommand{\arraystretch}{1}\begin{tabular}[c]{@{}#1@{}}#2\end{tabular}}}
\begin{tabular*}{\columnwidth}{@{\extracolsep{\fill}}llrr@{}}
\toprule
\textbf{Workload} & \textbf{Size} & \workloadcell[c]{\textbf{Input}\\\textbf{tokens}} & \workloadcell[c]{\textbf{Output}\\\textbf{tokens}} \\
\midrule
LooGLE & \workloadcell{105 documents\\1,951 requests} & 22,083--22,914 & 14--15 \\
\addlinespace[3pt]
NarrativeQA & \workloadcell{40 documents\\1,187 requests} & 52,797--54,618 & 6 \\
\addlinespace[3pt]
SWE-bench & \workloadcell{665 sessions\\23,356 turns} & 25,185--27,510 & 212 \\
\bottomrule
\end{tabular*}
\end{table}

%% file: figures/e2e-h200.tex
\begin{figure*}[!t]
\centering
\input{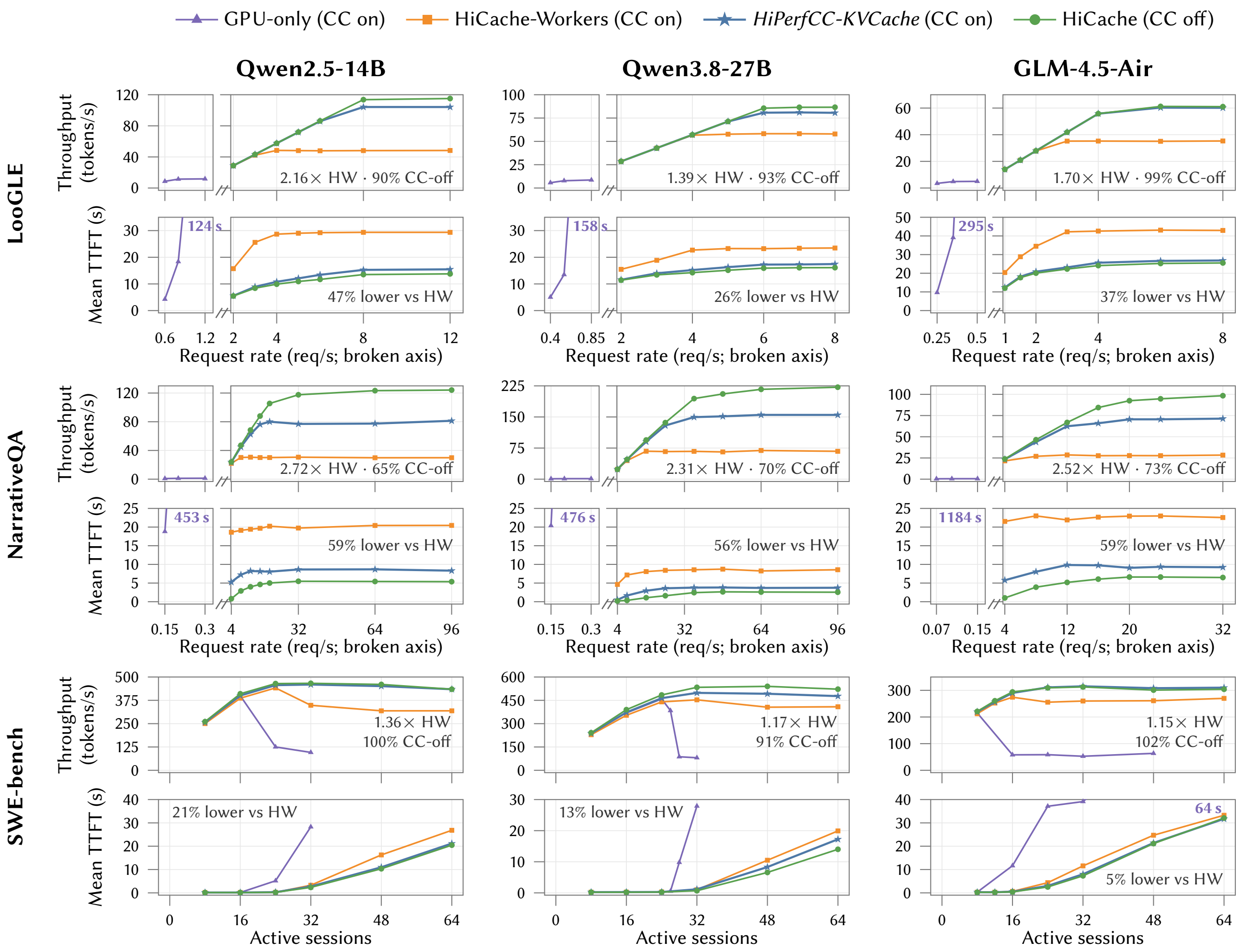}
\caption{Serving performance on H200.
Annotations show \sysname's gains over \baseline (\baselineshort) and
throughput relative to CC-off HiCache at peak load.
Purple labels report GPU-only TTFT at its highest measured
load when exceeding the plotted range.}
\label{fig:e2e-h200}
\end{figure*}

%% file: figures/gap-mechanisms-h200.tex
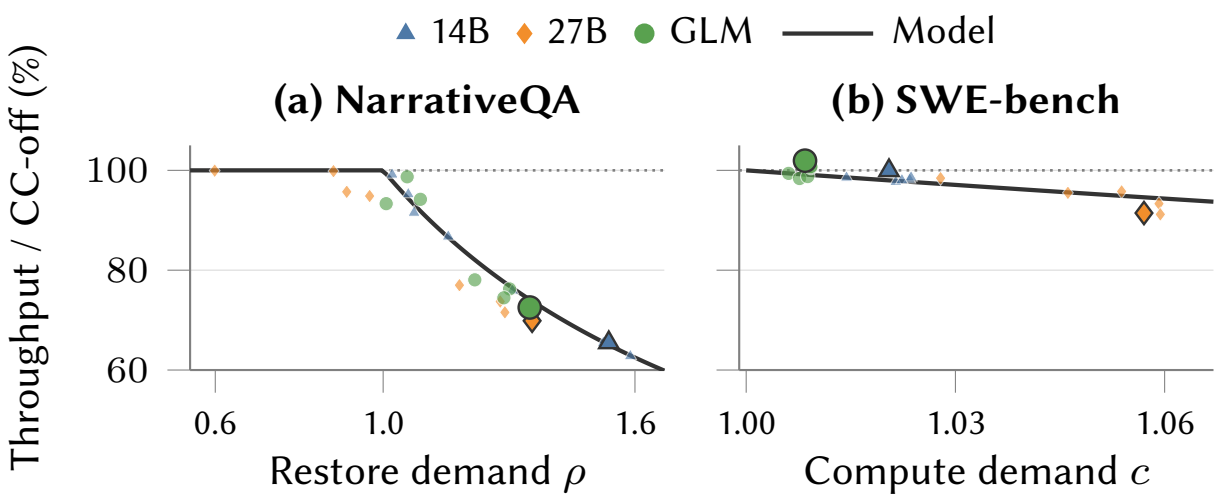
\begin{figure}[t]
\centering
\input{figures/gap-mechanisms-h200-plot}
\caption{Throughput relative to CC-off on NarrativeQA and SWE-bench.
Lines show model predictions: $1/\max(1,\rho)$ in (a) and $1/c$
in (b). Outlines mark each model's highest tested load.}
\label{fig:gap-mechanisms-h200}
\end{figure}

%% file: figures/gap-mechanisms-h200-plot.tex
\begingroup
\figurefont\selectfont
\newcommand{\mechanismseries}[4]{%
    \addplot[only marks,#2,mark=#3,mark size=1.35pt,opacity=.65,
        mark options={draw=white,line width=.2pt},forget plot,
        restrict expr to domain={\thisrow{series}+100*\thisrow{highest_load}}{#1:#1}]
        table[x=#4,y=measured_pct,col sep=comma,row sep=\\] {
series,model,workload,level,compute_demand,restore_demand,measured_pct,estimated_pct,highest_load\\
1,14B,narrativeqa,4,0.23123932,1.02145271,99.15878344,97.89978411,0\\
1,14B,narrativeqa,8,0.43197120,1.05990343,95.22703119,94.34821852,0\\
1,14B,narrativeqa,12,0.60360276,1.07365213,91.51682296,93.14003758,0\\
1,14B,narrativeqa,16,0.76207142,1.15519594,86.74996880,86.56540143,0\\
1,14B,narrativeqa,20,0.90549415,1.30792652,75.91306208,76.45689456,0\\
1,14B,narrativeqa,32,1.00245682,1.52466164,65.44852420,65.58832291,0\\
1,14B,narrativeqa,64,1.04973928,1.58901305,62.78508658,62.93214539,0\\
1,14B,narrativeqa,96,1.05621255,1.53760612,65.49345040,65.03616169,1\\
2,14B,swe,8,1.01438547,0.00100946,98.56868095,98.58185365,0\\
2,14B,swe,16,1.02145570,0.00072268,97.69008819,97.89949758,0\\
2,14B,swe,24,1.02341577,0.03019128,98.18473976,97.71199872,0\\
2,14B,swe,32,1.02364327,0.40278546,98.58456500,97.69028197,0\\
2,14B,swe,48,1.02233073,0.39525653,98.02309706,97.81570389,0\\
2,14B,swe,64,1.02047567,0.39978022,99.96317945,97.99351711,1\\
4,27B,narrativeqa,4,0.19836663,0.59862566,99.94165685,100.00000000,0\\
4,27B,narrativeqa,8,0.38506444,0.88164062,99.83434555,100.00000000,0\\
4,27B,narrativeqa,16,0.65661510,0.91326096,95.70253021,100.00000000,0\\
4,27B,narrativeqa,24,0.84671119,0.96782291,94.85490862,100.00000000,0\\
4,27B,narrativeqa,36,1.09088663,1.18178338,76.99163275,84.61787638,0\\
4,27B,narrativeqa,48,1.12532214,1.27895868,73.72484475,78.18860867,0\\
4,27B,narrativeqa,64,1.17405607,1.29008773,71.55299383,77.51410824,0\\
4,27B,narrativeqa,96,1.19898038,1.35496214,69.86957899,73.80280029,1\\
5,27B,swe,8,1.02787400,0.00020909,98.42964087,97.28818869,0\\
5,27B,swe,16,1.04613089,0.00047207,95.49427488,95.59033294,0\\
5,27B,swe,24,1.05383240,0.00058473,95.77299588,94.89174920,0\\
5,27B,swe,32,1.05918828,0.13875524,93.32179855,94.41192110,0\\
5,27B,swe,48,1.05938375,0.21238168,91.18275387,94.39450089,0\\
5,27B,swe,64,1.05703811,0.21628569,91.44937720,94.60396862,1\\
7,GLM,narrativeqa,4,0.26310645,1.05725592,98.71071407,94.58447859,0\\
7,GLM,narrativeqa,8,0.49613806,1.08882366,94.19691440,91.84223672,0\\
7,GLM,narrativeqa,12,0.70243504,1.00724790,93.32783665,99.28042558,0\\
7,GLM,narrativeqa,16,0.88115324,1.21836424,78.06750585,82.07726114,0\\
7,GLM,narrativeqa,20,0.96358143,1.30119197,76.24543862,76.85261073,0\\
7,GLM,narrativeqa,24,0.98640808,1.28761782,74.46771262,77.66279593,0\\
7,GLM,narrativeqa,32,1.02316715,1.34930152,72.52601970,74.11241951,1\\
8,GLM,swe,8,1.00608248,0.00091122,99.39274615,99.39542942,0\\
8,GLM,swe,12,1.00764459,0.00111715,98.36320176,99.24134044,0\\
8,GLM,swe,16,1.00882594,0.04340420,98.75173685,99.12512755,0\\
8,GLM,swe,24,1.00927505,0.24298327,100.56035690,99.08101881,0\\
8,GLM,swe,32,1.00927505,0.26900165,100.76843143,99.08101881,0\\
8,GLM,swe,48,1.00838660,0.27062855,102.39063244,99.16831542,0\\
8,GLM,swe,64,1.00843541,0.26852163,101.94079506,99.16351491,1\\
};%
    \addplot[only marks,#2,mark=#3,mark size=2.1pt,
        mark options={draw=VeloCCInk,line width=.45pt},forget plot,
        restrict expr to domain={\thisrow{series}+100*(1-\thisrow{highest_load})}{#1:#1}]
        table[x=#4,y=measured_pct,col sep=comma,row sep=\\] {
series,model,workload,level,compute_demand,restore_demand,measured_pct,estimated_pct,highest_load\\
1,14B,narrativeqa,4,0.23123932,1.02145271,99.15878344,97.89978411,0\\
1,14B,narrativeqa,8,0.43197120,1.05990343,95.22703119,94.34821852,0\\
1,14B,narrativeqa,12,0.60360276,1.07365213,91.51682296,93.14003758,0\\
1,14B,narrativeqa,16,0.76207142,1.15519594,86.74996880,86.56540143,0\\
1,14B,narrativeqa,20,0.90549415,1.30792652,75.91306208,76.45689456,0\\
1,14B,narrativeqa,32,1.00245682,1.52466164,65.44852420,65.58832291,0\\
1,14B,narrativeqa,64,1.04973928,1.58901305,62.78508658,62.93214539,0\\
1,14B,narrativeqa,96,1.05621255,1.53760612,65.49345040,65.03616169,1\\
2,14B,swe,8,1.01438547,0.00100946,98.56868095,98.58185365,0\\
2,14B,swe,16,1.02145570,0.00072268,97.69008819,97.89949758,0\\
2,14B,swe,24,1.02341577,0.03019128,98.18473976,97.71199872,0\\
2,14B,swe,32,1.02364327,0.40278546,98.58456500,97.69028197,0\\
2,14B,swe,48,1.02233073,0.39525653,98.02309706,97.81570389,0\\
2,14B,swe,64,1.02047567,0.39978022,99.96317945,97.99351711,1\\
4,27B,narrativeqa,4,0.19836663,0.59862566,99.94165685,100.00000000,0\\
4,27B,narrativeqa,8,0.38506444,0.88164062,99.83434555,100.00000000,0\\
4,27B,narrativeqa,16,0.65661510,0.91326096,95.70253021,100.00000000,0\\
4,27B,narrativeqa,24,0.84671119,0.96782291,94.85490862,100.00000000,0\\
4,27B,narrativeqa,36,1.09088663,1.18178338,76.99163275,84.61787638,0\\
4,27B,narrativeqa,48,1.12532214,1.27895868,73.72484475,78.18860867,0\\
4,27B,narrativeqa,64,1.17405607,1.29008773,71.55299383,77.51410824,0\\
4,27B,narrativeqa,96,1.19898038,1.35496214,69.86957899,73.80280029,1\\
5,27B,swe,8,1.02787400,0.00020909,98.42964087,97.28818869,0\\
5,27B,swe,16,1.04613089,0.00047207,95.49427488,95.59033294,0\\
5,27B,swe,24,1.05383240,0.00058473,95.77299588,94.89174920,0\\
5,27B,swe,32,1.05918828,0.13875524,93.32179855,94.41192110,0\\
5,27B,swe,48,1.05938375,0.21238168,91.18275387,94.39450089,0\\
5,27B,swe,64,1.05703811,0.21628569,91.44937720,94.60396862,1\\
7,GLM,narrativeqa,4,0.26310645,1.05725592,98.71071407,94.58447859,0\\
7,GLM,narrativeqa,8,0.49613806,1.08882366,94.19691440,91.84223672,0\\
7,GLM,narrativeqa,12,0.70243504,1.00724790,93.32783665,99.28042558,0\\
7,GLM,narrativeqa,16,0.88115324,1.21836424,78.06750585,82.07726114,0\\
7,GLM,narrativeqa,20,0.96358143,1.30119197,76.24543862,76.85261073,0\\
7,GLM,narrativeqa,24,0.98640808,1.28761782,74.46771262,77.66279593,0\\
7,GLM,narrativeqa,32,1.02316715,1.34930152,72.52601970,74.11241951,1\\
8,GLM,swe,8,1.00608248,0.00091122,99.39274615,99.39542942,0\\
8,GLM,swe,12,1.00764459,0.00111715,98.36320176,99.24134044,0\\
8,GLM,swe,16,1.00882594,0.04340420,98.75173685,99.12512755,0\\
8,GLM,swe,24,1.00927505,0.24298327,100.56035690,99.08101881,0\\
8,GLM,swe,32,1.00927505,0.26900165,100.76843143,99.08101881,0\\
8,GLM,swe,48,1.00838660,0.27062855,102.39063244,99.16831542,0\\
8,GLM,swe,64,1.00843541,0.26852163,101.94079506,99.16351491,1\\
};%
}
\begin{tikzpicture}
\begin{groupplot}[
    group style={group size=2 by 1,horizontal sep=.06\columnwidth},
    scale only axis,
    width=.38\columnwidth,height=.18\columnwidth,
    ymin=60,ymax=105,ytick={60,80,100},
    label style={font=\figurefont\fontsize{8}{9}\selectfont},
    tick label style={font=\figurefont\fontsize{7}{8}\selectfont},
    title style={font=\figurefont\fontsize{8}{9}\selectfont\bfseries,yshift=-5pt},
    xlabel style={yshift=3pt},
    tick align=outside,tick style={black!50},
    axis lines*=left,axis line style={black!50},
    ymajorgrids=true,grid style={black!12,line width=.25pt},
    scaled ticks=false,
    clip marker paths=true,
]
\nextgroupplot[
    xmin=.54,xmax=1.67,xtick={.6,1,1.6},xticklabels={0.6,1.0,1.6},
    title={(a) NarrativeQA},
    xlabel={Restore demand $\rho$},ylabel={Throughput / CC-off (\%)},
    legend columns=4,
    legend style={at={(1.08,1.32)},anchor=south,draw=none,
        font=\figurefont\fontsize{7.5}{8.5}\selectfont,
        column sep=1pt,/tikz/every even column/.append style={column sep=3pt}},
]
\addplot[black!50,densely dotted,line width=.5pt,forget plot]
    coordinates {(.54,100) (1.67,100)};
\addplot[VeloCCInk,line width=.8pt,domain=.54:1.67,samples=100,forget plot]
    {100/max(1,x)};
\mechanismseries{1}{VeloCCProtected}{triangle*}{restore_demand}
\mechanismseries{4}{VeloCCCipher}{diamond*}{restore_demand}
\mechanismseries{7}{VeloCCBackup}{*}{restore_demand}
\addlegendimage{only marks,VeloCCProtected,mark=triangle*,mark size=1.7pt}
\addlegendentry{14B}
\addlegendimage{only marks,VeloCCCipher,mark=diamond*,mark size=1.7pt}
\addlegendentry{27B}
\addlegendimage{only marks,VeloCCBackup,mark=*,mark size=1.7pt}
\addlegendentry{GLM}
\addlegendimage{VeloCCInk,line width=.8pt}
\addlegendentry{Model}

\nextgroupplot[
    xmin=.999,xmax=1.067,xtick={1,1.03,1.06},xticklabels={1.00,1.03,1.06},
    yticklabels=\empty,
    title={(b) SWE-bench},xlabel={Compute demand $c$},
]
\addplot[black!50,densely dotted,line width=.5pt,forget plot]
    coordinates {(.999,100) (1.067,100)};
\addplot[VeloCCInk,line width=.8pt,domain=1:1.067,samples=100,forget plot]
    {100/x};
\mechanismseries{2}{VeloCCProtected}{triangle*}{compute_demand}
\mechanismseries{5}{VeloCCCipher}{diamond*}{compute_demand}
\mechanismseries{8}{VeloCCBackup}{*}{compute_demand}
\end{groupplot}
\end{tikzpicture}
\endgroup

%% file: tables/ablation-h200.tex
\begin{table}[t]
\centering
\caption{Transfer-path ablation on H200 with Qwen2.5-14B on LooGLE (12 requests/s). Speedup is relative to \baseline. Full P3 is \sysname.}
\label{tab:ablation-h200}
\scriptsize
\setlength{\tabcolsep}{2pt}
\renewcommand{\arraystretch}{1.1}
\begin{tabular*}{\columnwidth}{@{\extracolsep{\fill}}lrrr@{}}
\toprule
\textbf{Configuration} & \textbf{Throughput} (tok/s) $\uparrow$ & \textbf{Speedup} $\uparrow$ & \textbf{Mean TTFT} (s) $\downarrow$ \\
\midrule
\baseline & 48.3 & $1.00\times$ & 29.4 \\
+ P1 & 63.3 & $1.31\times$ & 25.7 \\
+ P2 & 64.8 & $1.34\times$ & 25.1 \\
+ P3 base & 99.0 & $2.05\times$ & 16.3 \\
\quad + balancing only & 101.5 & $2.10\times$ & 15.8 \\
\quad + overlap only & 101.8 & $2.11\times$ & 15.9 \\
\textbf{+ full P3} & \textbf{105.3} & $\mathbf{2.18}\times$ & \textbf{15.2} \\
\midrule
HiCache (CC off) & 115.1 & $2.38\times$ & 13.8 \\
\bottomrule
\end{tabular*}
\par\vspace{3pt}
\parbox{\columnwidth}{\raggedright P1/P2 are cumulative. P3 variants retain P1+P2 and use a separate matched batch with three restore contexts, ordinary host memory, and VMM-shared GPU staging without an assembly copy. P3 base disables balancing and H2D--scatter overlap.}
\end{table}

%% file: tables/e2e-pro6000.tex
\begin{table}[t]
\centering
\caption{Serving on RTX PRO 6000 Blackwell. \baselineshort: \baseline; Ours: \sysname; CC-off: HiCache without CC.}
\label{tab:e2e-pro6000}
\fontsize{7.5}{9}\selectfont
\setlength{\tabcolsep}{1.5pt}
\renewcommand{\arraystretch}{1.12}
\begin{tabular*}{\columnwidth}{@{\extracolsep{\fill}}llrrr@{\hspace{5pt}}rrr@{}}
\toprule
 & & \multicolumn{3}{c}{\textbf{Throughput} (tok/s) $\uparrow$} & \multicolumn{3}{c}{\textbf{Mean TTFT} (s) $\downarrow$} \\
\cmidrule(lr){3-5}\cmidrule(l){6-8}
\textbf{Model} & \textbf{Workload} & \baselineshort & \textbf{Ours} & CC-off & \baselineshort & \textbf{Ours} & CC-off \\
\midrule
14B & LooGLE & 16.2 & \textbf{21.2} & 21.0 & 100.8 & \textbf{78.0} & 77.6 \\
 & NarrativeQA & 45.0 & \textbf{99.1} & 110.9 & 13.8 & \textbf{5.8} & 5.0 \\
 & SWE-bench & 197.2 & \textbf{210.6} & 211.2 & 52.2 & \textbf{48.0} & 46.8 \\
\addlinespace[2pt]
27B & LooGLE & 45.3 & \textbf{50.0} & 51.3 & 36.3 & \textbf{33.1} & 32.0 \\
 & NarrativeQA & 74.9 & \textbf{83.4} & 107.7 & 8.7 & \textbf{7.9} & 5.8 \\
 & SWE-bench & 205.2 & \textbf{203.8} & 212.7 & 47.3 & \textbf{49.6} & 48.3 \\
\addlinespace[2pt]
GLM & LooGLE & 13.6 & \textbf{17.5} & 17.1 & 125.8 & \textbf{98.0} & 99.7 \\
 & NarrativeQA & 19.7 & \textbf{35.6} & 63.5 & 36.1 & \textbf{19.1} & 10.8 \\
 & SWE-bench & 151.7 & \textbf{158.7} & 162.5 & 72.4 & \textbf{71.4} & 69.6 \\
\bottomrule
\end{tabular*}
\end{table}

%% file: sections/07-related-work.tex
\section{Related Work}
\label{sec:related}

\bheading{Confidential GPU serving.}
Prior studies identify transfer and framework overheads in
confidential GPU inference~\cite{ispass_cc,blackwell}.
Background output workers recover overlap between token
readback and computation~\cite{blackwell,serializedbridge}.
Serialized Bridge evaluates synchronous scheduling for KV churn
and uses context parallelism to accelerate
model loading~\cite{serializedbridge}.
\sysname identifies KV-transfer driver-lock and allocation
constraints, then coordinates resource lifetimes, backups,
and parallel restoration while preserving overlap scheduling.

\bheading{Reducing protected-transfer costs.}
PipeLLM predicts transfers and initialization vectors to
pre-encrypt data~\cite{pipellm}.
\laegis proposes GPU extensions for flexible initialization-vector
management, enabling page pre-encryption during unified-memory
migration~\cite{laegis}.
Fastrack combines a direct client--GPU secure channel,
parallel authentication, and pipelined cryptography~\cite{fastrack}.
\sysname reduces interference between protected KV transfers
and foreground serving via runtime coordination with existing CUDA APIs.

%% file: sections/09-conclusion.tex
\section{Conclusion}
\label{sec:conclusion}

Hierarchical KV caching can lose its benefit under GPU CC
when protected transfers stall independent serving work.
We identify the driver-lock conflicts behind these stalls
and the allocation constraints on transfer parallelism.
Guided by these findings, \sysname coordinates resource
lifetimes, backup submissions, and parallel restoration
using existing protected CUDA interfaces.
Implemented in SGLang, \sysname achieves a geometric-mean
throughput speedup of $1.74\times$ over \baseline on H200
at peak load, with gains also
observed on Blackwell.

%% file: sections/10-appendix.tex
\section{Motivation Experiment Details}
\label{sec:appendix:motivation}

\bheading{Serving configurations.}
The SGLang experiments in \secref{sec:motivation} use the same
software stack as the main evaluation.
Across CC modes, all configurations use a GPU
KV-cache capacity of 163,840 tokens and a host-cache capacity
of 500~GB whenever host caching is enabled.
Under CC, the host cache uses CUDA host allocations
through PyTorch's pinned-memory allocator in place of the unsupported
\texttt{cudaHostRegister} path~\cite{hoppercc}.

\bheading{Workload and measurement.}
The workload contains 40 conversation sessions with eight turns each,
drawing initial passages from LooGLE.
Each session starts with 8,192 input tokens.
Follow-up requests retain the generated conversation history and
add 256 input tokens, and each response generates 256 tokens.
The client maintains eight concurrent requests in closed loop
with no added think time.
Before each measured run, we issue eight warm-up requests and then
flush the GPU cache and any enabled host cache.
\tabref{tab:motivation-stock-backends} reports the results
for each configuration and CC mode.

\input{tables/motivation-stock-backends}

\section{Driver-Lock Measurements}
\label{sec:appendix:characterization}

\input{tables/cuda-api-lock-details}

\bheading{Lock probes.}
On H200 and RTX PRO 6000 under CC, a background thread issues
pinned H2D or D2H copies through \texttt{cudaMemcpyAsync} or
\texttt{cudaMemcpyBatchAsync}.
Foreground operations run on the copying stream, another stream
in the same context, or a separate prewarmed context.
We match foreground waits to background holds by lock address
within each process, identifying the same C, S, and T instances
for both copy APIs.
Latency measurements use separate, uninstrumented runs and idle
controls.

Single-buffer probes use 1~GiB background copies and the foreground
APIs in \tabref{tab:cuda-api-lock-details}, invoked 10~ms later.
Event query/wait probes use completed events, and asynchronous
allocation/free uses a warm pool.
Batch probes test event creation, event recording, kernel launches,
and small copies.
CC-off controls cover both GPUs for single-buffer calls and H200
for batches.

\bheading{Foreground progress within a batch.}
For the configurations in \tabref{tab:batch-copy-progress},
foreground copies arrive at uniformly sampled offsets after the
background call starts. All begin before it returns.
Each configuration spans two fresh processes.
Completion includes foreground stream synchronization.
Only 2.5--22.5\% of foreground copies complete before the batch
returns, despite T being released between buffers.

\input{tables/batch-copy-progress}

\section{Design Experiment Details}
\label{sec:appendix:design}

All measurements below use H200 with CC enabled.

\bheading{Copy-granularity measurements.}
In \figref{fig:copy-granularity}, bandwidth includes final stream
synchronization. Host-call duration measures only time in the copy API.
Each configuration has five measured samples per run.

\bheading{Foreground-readback probe.}
A backup worker transfers 1.5~GiB through consecutive 6~MiB
D2H submissions, synchronizing after each.
A second thread issues a 32-byte pinned D2H readback at a random
offset after backup starts, on another stream in the same context.
Completion latency includes stream synchronization.
We pool 90 trials from three fresh processes.

\bheading{Ordinary-memory validation.}
We repeat these D2H probes with ordinary host memory as the bulk
copy destination, keeping the readback buffer pinned.
With 6~MiB buffers, reducing bytes per call from 768 to 6~MiB
preserves approximately 8.2~GB/s bandwidth while reducing median
host-call duration from about 99 to 0.74~ms.
With 90 trials per condition, the priority gate reduces
90th-percentile readback latency from about 129 to 0.78~ms,
including gate waiting and stream synchronization.

\bheading{Context-parallelism measurements.}
Each worker uses a separate CPU core, CUDA context, and nonblocking stream.
GPU buffers are shared through CUDA's virtual-memory interfaces,
with disjoint H2D and D2H regions.
The pinned host pool uses default \texttt{cudaHostAlloc} flags.
\figref{fig:context-parallelism} reports bandwidth by counting completed
bytes over a common 20-second window after warmup.

\section{Evaluation Details}
\label{sec:appendix:evaluation}

\bheading{Blackwell setup.}
The RTX PRO 6000 experiments in \secref{sec:evaluation:generalization}
retain the implementation, serving configurations, and measurement
rules of \secref{sec:evaluation:setup}, using FlashInfer attention
and a 330~GB host cache.
NarrativeQA prepopulates the host cache with KV for 30 documents
to fit its capacity.

\subsection{Inter-Token Latency}
\label{sec:appendix:itl}

\input{figures/itl-h200}

\figref{fig:itl-h200} reports mean client-observed ITL for the H200
evaluation.
We pool consecutive-token intervals across the same requests used
for TTFT in \secref{sec:evaluation:setup}.

\bheading{Comparison with \baseline.}
At each model--workload pair's peak load,
the relative reductions in mean ITL average 55\% across the nine
pairs and reach up to 95\% over \baseline.
These gains are consistent with reduced driver-lock stalls
(\secref{sec:characterization:locks}).

\bheading{Comparison with CC-off.}
At the same peak loads, relative changes in mean ITL range
from a 12\% increase to a 51\% reduction compared with native
CC-off HiCache.
Analysis of serving counters in the document runs with lower ITL
suggests that slower protected restores space out request admissions,
leaving smaller decode batches and fewer interleaved prefills.

\subsection{Throughput Model}
\label{sec:appendix:gap-model}

\bheading{Phase accounting.}
We split each LooGLE run into initial population and reuse when
every document has produced its first response token.
CC-off phases define the durations $T_p$ and weights
$f_p=T_p/\sum_q T_q$ in Equation~\ref{eq:cc-gap-model}.
Corresponding CC-on phases determine the restoration work
assigned to each phase.
NarrativeQA and SWE-bench each use one phase covering the
evaluation interval defined in \secref{sec:evaluation:setup}.

\bheading{Compute demand.}
For model $m$, let $u$ be a proxy for baseline compute demand,
$\delta_m$ the additional CC time per prefill batch, and $n_p$
the CC-off prefill-batch rate in phase $p$.
We estimate $c_p=u+\delta_m n_p$.
For document workloads, $u$ is CC-off request throughput divided
by the maximum on the same model--workload curve.
For SWE-bench, $u=1$ approximates a busy closed-loop server.
We calibrate $\delta_m$ from the CC-on/off difference in mean
per-request prefill-forward latency at eight SWE-bench sessions:
10.50, 23.46, and 5.86~ms for 14B, 27B, and GLM, respectively.
Because 97--98\% of these batches contain one request, we
approximate this difference as additional time per prefill batch.

For 27B, a 64-token prefill without KV transfers launches roughly
1,000 linear-attention kernels outside CUDA graphs.
Unprofiled short-prefill probes show about 17~ms of additional
CC latency, supporting the launch-cost explanation in
\secref{sec:evaluation:gap}.

\bheading{Restoration and admission waits.}
We allocate measured CC-on whole-run bytes and restore counts to
phases in proportion to host-hit tokens and requests with host
hits, respectively.
Scaling each phase's totals by its CC-off/CC-on request-count
ratio and dividing by $T_p$ yields the restored-byte rate $D_p$
and restore-operation rate $r_p$.
Byte counts include the hybrid model's recurrent state.

A completed restore retains its slot until owner admission,
incurring an additional mean wait $g$ after copy completion.
We estimate $\rho_p=D_p/B+r_p g$, using $B=18.9$~GB/s measured
over active restoration time.
We approximate $g$ as half the paired CC-off median ITL, assuming
periodic admission opportunities and uniformly phased completions.

\input{figures/gap-model-h200}

\bheading{Comparison.}
\figref{fig:gap-model-h200} compares the model's reconstruction
from measured inputs with all 63 paired configurations.
With $B$, per-model $\delta_m$, and the half-ITL rule fixed across
loads, the mean absolute error is 1.51 percentage points and the
maximum is 7.63 points.
The comparison uses measured CC-on restoration work and does not
constitute held-out validation.

%% file: tables/motivation-stock-backends.tex
\begin{table}[t]
\centering
\caption{Output-token throughput and mean TTFT for GPU-only
serving and stock HiCache backends on the H200 motivation workload.}
\label{tab:motivation-stock-backends}
\small
\setlength{\tabcolsep}{3pt}
\renewcommand{\arraystretch}{1.1}
\begin{tabular*}{\columnwidth}{@{\extracolsep{\fill}}lrrrr@{}}
\toprule
\textbf{Configuration}
& \multicolumn{2}{c}{\textbf{Throughput} (tok/s) $\uparrow$}
& \multicolumn{2}{c}{\textbf{Mean TTFT} (s) $\downarrow$} \\
\cmidrule(lr){2-3}
\cmidrule(lr){4-5}
& \textbf{CC off} & \textbf{CC on}
& \textbf{CC off} & \textbf{CC on} \\
\midrule
GPU-only & 317 & 314 & 1.69 & 1.71 \\
HiCache-kernel & 474 & 152 & 0.52 & 9.15 \\
HiCache-direct & 470 & 269 & 0.61 & 3.39 \\
\bottomrule
\end{tabular*}
\end{table}

%% file: tables/cuda-api-lock-details.tex
\begin{table}[t]
\centering
\caption{Foreground API variants tested against bulk
\texttt{cudaMemcpyAsync} under CC on H200 and RTX PRO 6000.
Lock notation follows \tabref{tab:cuda-api-locks}.
Copy sizes are 4~KiB unless specified.
``None'' indicates no observed acquisition of C, S, or T.
The pageable-H2D threshold persists with fixed source and
destination allocations, indicating that it depends on copy length.}
\label{tab:cuda-api-lock-details}
\small
\setlength{\tabcolsep}{4pt}
\renewcommand{\arraystretch}{1.1}
\begin{tabular*}{\columnwidth}
  {@{\extracolsep{\fill}}p{0.70\columnwidth}l@{}}
\toprule
\textbf{API variants} & \textbf{Result} \\
\midrule
\texttt{cudaEventCreateWithFlags},
\texttt{cudaEventDestroy},
\texttt{cudaStreamCreateWithFlags},
\texttt{cudaStreamDestroy},
\texttt{cudaHostAlloc},
\texttt{cudaFreeHost},
\texttt{cudaMalloc},
\texttt{cudaFree}
& $C_{\mathrm{W}}$ \\
\addlinespace

Kernel launch,
\texttt{cudaGraphLaunch},
\texttt{cudaEventRecord}
& $C_{\mathrm{R}} + S$ \\
\addlinespace

\texttt{cudaMemcpyAsync}: pinned H2D,
pinned/pageable D2H
& $C_{\mathrm{R}} + S + T$ \\
\addlinespace

\texttt{cudaMemcpyAsync}: pageable H2D,
copy size $\leq64$~KiB
& $C_{\mathrm{R}} + S$ \\
\addlinespace

\texttt{cudaMemcpyAsync}: pageable H2D,
copy size $>64$~KiB
& $C_{\mathrm{R}} + S + T$ \\
\addlinespace

\texttt{cudaEventQuery},
\texttt{cudaEventSynchronize},
\texttt{cudaStreamQuery},
\texttt{cudaStreamSynchronize},
\texttt{cudaStreamWaitEvent},
\texttt{cudaMemGetInfo},
\texttt{cudaPointerGetAttributes}
& None \\
\addlinespace

\texttt{cudaMallocAsync},
\texttt{cudaFreeAsync}
& None \\
\addlinespace

\texttt{cudaHostRegister},
\texttt{cudaHostUnregister}
& Unsupported \\
\bottomrule
\end{tabular*}
\end{table}

%% file: tables/batch-copy-progress.tex
\begin{table}[t]
\centering
\caption{Foreground completion before batch API return on H200
under CC. A 32-byte pinned copy competes with 6~MiB batch buffers in
the same direction on another stream in the same context.
Entries report counts and percentages over 40 uninstrumented trials.}
\label{tab:batch-copy-progress}
\small
\setlength{\tabcolsep}{4pt}
\renewcommand{\arraystretch}{1.1}
\begin{tabular*}{\columnwidth}{@{\extracolsep{\fill}}lcc@{}}
\toprule
\textbf{Bytes per batch} & \textbf{H2D} & \textbf{D2H} \\
\midrule
24~MiB & 2/40 (5\%) & 1/40 (2.5\%) \\
768~MiB & 5/40 (12.5\%) & 9/40 (22.5\%) \\
\bottomrule
\end{tabular*}
\end{table}

%% file: figures/itl-h200.tex
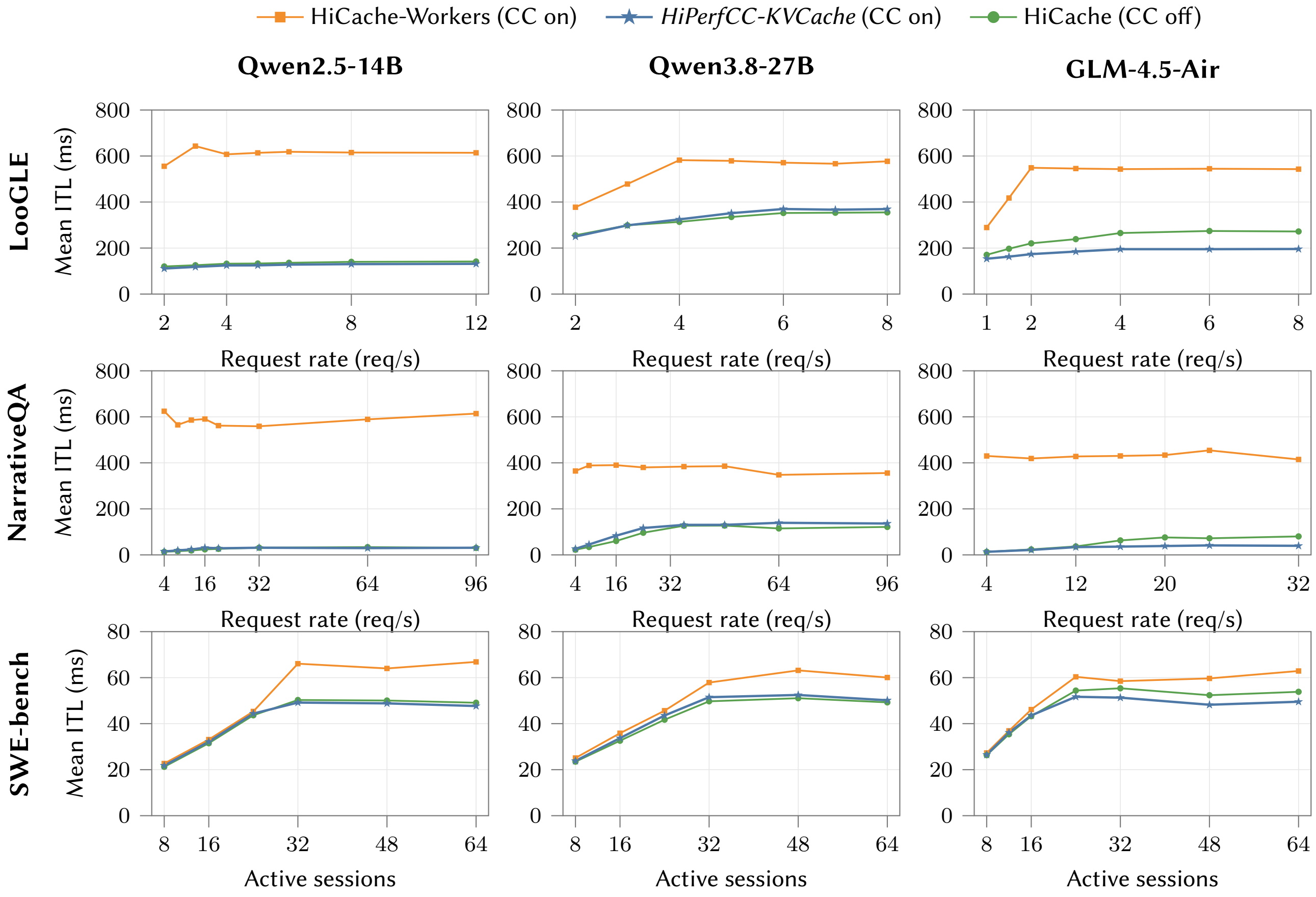
\begin{figure*}[!t]
\centering
\input{figures/itl-h200-grid}
\caption{Mean inter-token latency on H200 (lower is better).
Curves show all measured loads for the three host-cache configurations.}
\label{fig:itl-h200}
\end{figure*}

%% file: figures/itl-h200-grid.tex
\resizebox{\textwidth}{!}{%
\begin{tikzpicture}[font=\figurefont\fontsize{9}{10}\selectfont]
\begin{groupplot}[scale only axis,width=4.30cm,height=2.35cm,
group style={group size=3 by 3,horizontal sep=.95cm,vertical sep=.98cm},
axis line style={VeloCCInk!65,line width=.4pt},
tick style={VeloCCInk!65,line width=.4pt},tick align=outside,tick pos=left,
minor tick length=0pt,scaled ticks=false,
tick label style={font=\figurefont\fontsize{8}{9}\selectfont},
label style={font=\figurefont\fontsize{9}{10}\selectfont},
title style={font=\figurefont\bfseries\fontsize{10}{11}\selectfont,yshift=2pt},
grid=major,grid style={VeloCCInk!12,line width=.3pt},clip=true]
\nextgroupplot[xmin=1.6,xmax=12.4,ymin=0,ymax=800,ytick={0,200,400,600,800},xtick={2,4,8,12},xlabel={Request rate (req/s)},ylabel={Mean ITL (ms)},title={Qwen2.5-14B}]
\addplot[forget plot,e2e background] table[x=level,y=background_ms,col sep=comma,row sep=\\] {
level,background_ms,hicache_ccoff_ms,velocc_ms\\
2,555.5010901369022,119.97763587953227,110.94839277773325\\
3,643.1065745545656,125.33071896402426,118.06971655318044\\
4,607.384225647354,131.99386871054506,124.38792454098918\\
5,613.8183502633939,132.77872563342612,124.75159806871473\\
6,618.381471316517,135.89444369179597,127.89323785071655\\
8,615.1659112274181,140.0411118692179,130.0429206442288\\
12,614.0704292229557,141.45625920631574,131.36189271479725\\
};
\addplot[forget plot,e2e cc off] table[x=level,y=hicache_ccoff_ms,col sep=comma,row sep=\\] {
level,background_ms,hicache_ccoff_ms,velocc_ms\\
2,555.5010901369022,119.97763587953227,110.94839277773325\\
3,643.1065745545656,125.33071896402426,118.06971655318044\\
4,607.384225647354,131.99386871054506,124.38792454098918\\
5,613.8183502633939,132.77872563342612,124.75159806871473\\
6,618.381471316517,135.89444369179597,127.89323785071655\\
8,615.1659112274181,140.0411118692179,130.0429206442288\\
12,614.0704292229557,141.45625920631574,131.36189271479725\\
};
\addplot[forget plot,e2e velocc] table[x=level,y=velocc_ms,col sep=comma,row sep=\\] {
level,background_ms,hicache_ccoff_ms,velocc_ms\\
2,555.5010901369022,119.97763587953227,110.94839277773325\\
3,643.1065745545656,125.33071896402426,118.06971655318044\\
4,607.384225647354,131.99386871054506,124.38792454098918\\
5,613.8183502633939,132.77872563342612,124.75159806871473\\
6,618.381471316517,135.89444369179597,127.89323785071655\\
8,615.1659112274181,140.0411118692179,130.0429206442288\\
12,614.0704292229557,141.45625920631574,131.36189271479725\\
};
\nextgroupplot[xmin=1.76,xmax=8.24,ymin=0,ymax=800,ytick={0,200,400,600,800},xtick={2,4,6,8},xlabel={Request rate (req/s)},title={Qwen3.8-27B}]
\addplot[forget plot,e2e background] table[x=level,y=background_ms,col sep=comma,row sep=\\] {
level,background_ms,hicache_ccoff_ms,velocc_ms\\
2,377.47741797013356,255.63258247387827,250.15895723870761\\
3,478.2562551345009,298.9087919868345,298.6681460032481\\
4,582.1206816558515,313.73791906168526,324.6242026740087\\
5,578.9882158251789,335.02801985329495,351.6502652487388\\
6,571.0748842438411,352.3664513254631,369.699447607638\\
7,566.5171932059324,353.4781216202219,366.667452506619\\
8,577.0036402387033,355.02563615624183,369.3720141776866\\
};
\addplot[forget plot,e2e cc off] table[x=level,y=hicache_ccoff_ms,col sep=comma,row sep=\\] {
level,background_ms,hicache_ccoff_ms,velocc_ms\\
2,377.47741797013356,255.63258247387827,250.15895723870761\\
3,478.2562551345009,298.9087919868345,298.6681460032481\\
4,582.1206816558515,313.73791906168526,324.6242026740087\\
5,578.9882158251789,335.02801985329495,351.6502652487388\\
6,571.0748842438411,352.3664513254631,369.699447607638\\
7,566.5171932059324,353.4781216202219,366.667452506619\\
8,577.0036402387033,355.02563615624183,369.3720141776866\\
};
\addplot[forget plot,e2e velocc] table[x=level,y=velocc_ms,col sep=comma,row sep=\\] {
level,background_ms,hicache_ccoff_ms,velocc_ms\\
2,377.47741797013356,255.63258247387827,250.15895723870761\\
3,478.2562551345009,298.9087919868345,298.6681460032481\\
4,582.1206816558515,313.73791906168526,324.6242026740087\\
5,578.9882158251789,335.02801985329495,351.6502652487388\\
6,571.0748842438411,352.3664513254631,369.699447607638\\
7,566.5171932059324,353.4781216202219,366.667452506619\\
8,577.0036402387033,355.02563615624183,369.3720141776866\\
};
\nextgroupplot[xmin=0.72,xmax=8.28,ymin=0,ymax=800,ytick={0,200,400,600,800},xtick={1,2,4,6,8},xlabel={Request rate (req/s)},title={GLM-4.5-Air}]
\addplot[forget plot,e2e background] table[x=level,y=background_ms,col sep=comma,row sep=\\] {
level,background_ms,hicache_ccoff_ms,velocc_ms\\
1,289.2109266520851,170.4807471083403,153.482084537067\\
1.5,417.43019314071364,196.9548975776987,162.5827368138467\\
2,548.8623574047256,220.41905805854492,173.60079847528849\\
3,545.5051340267845,238.73180141964073,184.7630821308101\\
4,543.0129638665305,265.20371975105974,195.0094266712037\\
6,544.7056247478968,274.13706118041154,195.00481744785\\
8,542.9178516868868,272.0027593538118,195.86765561335162\\
};
\addplot[forget plot,e2e cc off] table[x=level,y=hicache_ccoff_ms,col sep=comma,row sep=\\] {
level,background_ms,hicache_ccoff_ms,velocc_ms\\
1,289.2109266520851,170.4807471083403,153.482084537067\\
1.5,417.43019314071364,196.9548975776987,162.5827368138467\\
2,548.8623574047256,220.41905805854492,173.60079847528849\\
3,545.5051340267845,238.73180141964073,184.7630821308101\\
4,543.0129638665305,265.20371975105974,195.0094266712037\\
6,544.7056247478968,274.13706118041154,195.00481744785\\
8,542.9178516868868,272.0027593538118,195.86765561335162\\
};
\addplot[forget plot,e2e velocc] table[x=level,y=velocc_ms,col sep=comma,row sep=\\] {
level,background_ms,hicache_ccoff_ms,velocc_ms\\
1,289.2109266520851,170.4807471083403,153.482084537067\\
1.5,417.43019314071364,196.9548975776987,162.5827368138467\\
2,548.8623574047256,220.41905805854492,173.60079847528849\\
3,545.5051340267845,238.73180141964073,184.7630821308101\\
4,543.0129638665305,265.20371975105974,195.0094266712037\\
6,544.7056247478968,274.13706118041154,195.00481744785\\
8,542.9178516868868,272.0027593538118,195.86765561335162\\
};
\nextgroupplot[xmin=0.32,xmax=99.68,ymin=0,ymax=800,ytick={0,200,400,600,800},xtick={4,16,32,64,96},xlabel={Request rate (req/s)},ylabel={Mean ITL (ms)}]
\addplot[forget plot,e2e background] table[x=level,y=background_ms,col sep=comma,row sep=\\] {
level,background_ms,hicache_ccoff_ms,velocc_ms\\
4,624.3779014416541,13.266769767144998,14.792829831459168\\
8,565.0393461618313,16.129180285436988,19.482331241231876\\
12,585.9441244424975,19.024216700690403,23.546751072520095\\
16,590.5403255835209,23.832366483305115,31.12316465990387\\
20,561.8058385141879,26.07573321517542,28.71474450312256\\
32,558.9445193857762,31.23938842285073,31.037473153611366\\
64,589.0474342820024,33.67796564512319,29.412812229534897\\
96,614.0717440991351,30.352869988752104,30.93907012198101\\
};
\addplot[forget plot,e2e cc off] table[x=level,y=hicache_ccoff_ms,col sep=comma,row sep=\\] {
level,background_ms,hicache_ccoff_ms,velocc_ms\\
4,624.3779014416541,13.266769767144998,14.792829831459168\\
8,565.0393461618313,16.129180285436988,19.482331241231876\\
12,585.9441244424975,19.024216700690403,23.546751072520095\\
16,590.5403255835209,23.832366483305115,31.12316465990387\\
20,561.8058385141879,26.07573321517542,28.71474450312256\\
32,558.9445193857762,31.23938842285073,31.037473153611366\\
64,589.0474342820024,33.67796564512319,29.412812229534897\\
96,614.0717440991351,30.352869988752104,30.93907012198101\\
};
\addplot[forget plot,e2e velocc] table[x=level,y=velocc_ms,col sep=comma,row sep=\\] {
level,background_ms,hicache_ccoff_ms,velocc_ms\\
4,624.3779014416541,13.266769767144998,14.792829831459168\\
8,565.0393461618313,16.129180285436988,19.482331241231876\\
12,585.9441244424975,19.024216700690403,23.546751072520095\\
16,590.5403255835209,23.832366483305115,31.12316465990387\\
20,561.8058385141879,26.07573321517542,28.71474450312256\\
32,558.9445193857762,31.23938842285073,31.037473153611366\\
64,589.0474342820024,33.67796564512319,29.412812229534897\\
96,614.0717440991351,30.352869988752104,30.93907012198101\\
};
\nextgroupplot[xmin=0.32,xmax=99.68,ymin=0,ymax=800,ytick={0,200,400,600,800},xtick={4,16,32,64,96},xlabel={Request rate (req/s)}]
\addplot[forget plot,e2e background] table[x=level,y=background_ms,col sep=comma,row sep=\\] {
level,background_ms,hicache_ccoff_ms,velocc_ms\\
4,364.5996580972192,22.466895405966387,26.26500831135576\\
8,388.50049543753374,34.04090227429755,45.08356386988077\\
16,390.0234055543701,60.25574394394704,83.4294074664262\\
24,380.05364109791327,96.19969379083263,116.78736953655208\\
36,383.7130989285716,126.63149384333056,130.4386778694943\\
48,385.6088456735178,127.19858183710542,130.5003407204878\\
64,347.7363855925483,114.9197162243393,139.3869407341119\\
96,355.72819244686633,121.41480261107108,136.2712202142349\\
};
\addplot[forget plot,e2e cc off] table[x=level,y=hicache_ccoff_ms,col sep=comma,row sep=\\] {
level,background_ms,hicache_ccoff_ms,velocc_ms\\
4,364.5996580972192,22.466895405966387,26.26500831135576\\
8,388.50049543753374,34.04090227429755,45.08356386988077\\
16,390.0234055543701,60.25574394394704,83.4294074664262\\
24,380.05364109791327,96.19969379083263,116.78736953655208\\
36,383.7130989285716,126.63149384333056,130.4386778694943\\
48,385.6088456735178,127.19858183710542,130.5003407204878\\
64,347.7363855925483,114.9197162243393,139.3869407341119\\
96,355.72819244686633,121.41480261107108,136.2712202142349\\
};
\addplot[forget plot,e2e velocc] table[x=level,y=velocc_ms,col sep=comma,row sep=\\] {
level,background_ms,hicache_ccoff_ms,velocc_ms\\
4,364.5996580972192,22.466895405966387,26.26500831135576\\
8,388.50049543753374,34.04090227429755,45.08356386988077\\
16,390.0234055543701,60.25574394394704,83.4294074664262\\
24,380.05364109791327,96.19969379083263,116.78736953655208\\
36,383.7130989285716,126.63149384333056,130.4386778694943\\
48,385.6088456735178,127.19858183710542,130.5003407204878\\
64,347.7363855925483,114.9197162243393,139.3869407341119\\
96,355.72819244686633,121.41480261107108,136.2712202142349\\
};
\nextgroupplot[xmin=2.88,xmax=33.12,ymin=0,ymax=800,ytick={0,200,400,600,800},xtick={4,12,20,32},xlabel={Request rate (req/s)}]
\addplot[forget plot,e2e background] table[x=level,y=background_ms,col sep=comma,row sep=\\] {
level,background_ms,hicache_ccoff_ms,velocc_ms\\
4,429.87743015851333,13.214987904571242,13.251923576403874\\
8,418.9649699328737,23.723858529874587,21.415443406752463\\
12,427.81443625547416,36.719319737597274,33.738340701842404\\
16,430.09998888836185,62.80816335480662,35.882959029176355\\
20,433.68146440623525,75.8697330271223,38.53627225954119\\
24,454.2540819672851,72.08619610490433,41.08977179644961\\
32,414.71356685143473,80.20286611635862,39.417686432540364\\
};
\addplot[forget plot,e2e cc off] table[x=level,y=hicache_ccoff_ms,col sep=comma,row sep=\\] {
level,background_ms,hicache_ccoff_ms,velocc_ms\\
4,429.87743015851333,13.214987904571242,13.251923576403874\\
8,418.9649699328737,23.723858529874587,21.415443406752463\\
12,427.81443625547416,36.719319737597274,33.738340701842404\\
16,430.09998888836185,62.80816335480662,35.882959029176355\\
20,433.68146440623525,75.8697330271223,38.53627225954119\\
24,454.2540819672851,72.08619610490433,41.08977179644961\\
32,414.71356685143473,80.20286611635862,39.417686432540364\\
};
\addplot[forget plot,e2e velocc] table[x=level,y=velocc_ms,col sep=comma,row sep=\\] {
level,background_ms,hicache_ccoff_ms,velocc_ms\\
4,429.87743015851333,13.214987904571242,13.251923576403874\\
8,418.9649699328737,23.723858529874587,21.415443406752463\\
12,427.81443625547416,36.719319737597274,33.738340701842404\\
16,430.09998888836185,62.80816335480662,35.882959029176355\\
20,433.68146440623525,75.8697330271223,38.53627225954119\\
24,454.2540819672851,72.08619610490433,41.08977179644961\\
32,414.71356685143473,80.20286611635862,39.417686432540364\\
};
\nextgroupplot[xmin=5.76,xmax=66.24,ymin=0,ymax=80,ytick={0,20,40,60,80},xtick={8,16,32,48,64},xlabel={Active sessions},ylabel={Mean ITL (ms)}]
\addplot[forget plot,e2e background] table[x=level,y=background_ms,col sep=comma,row sep=\\] {
level,background_ms,hicache_ccoff_ms,velocc_ms\\
8,22.679579532743436,21.24718574547014,21.772683614381478\\
16,33.09091863481983,31.496426239096273,32.14253033715899\\
24,45.307042015196444,43.640893470196644,44.380770344636524\\
32,66.06244024954047,50.281284522797776,49.162105058614536\\
48,63.99956228781187,50.05435580260274,48.81611749269989\\
64,66.85390773645142,49.08609383555707,47.66504717130042\\
};
\addplot[forget plot,e2e cc off] table[x=level,y=hicache_ccoff_ms,col sep=comma,row sep=\\] {
level,background_ms,hicache_ccoff_ms,velocc_ms\\
8,22.679579532743436,21.24718574547014,21.772683614381478\\
16,33.09091863481983,31.496426239096273,32.14253033715899\\
24,45.307042015196444,43.640893470196644,44.380770344636524\\
32,66.06244024954047,50.281284522797776,49.162105058614536\\
48,63.99956228781187,50.05435580260274,48.81611749269989\\
64,66.85390773645142,49.08609383555707,47.66504717130042\\
};
\addplot[forget plot,e2e velocc] table[x=level,y=velocc_ms,col sep=comma,row sep=\\] {
level,background_ms,hicache_ccoff_ms,velocc_ms\\
8,22.679579532743436,21.24718574547014,21.772683614381478\\
16,33.09091863481983,31.496426239096273,32.14253033715899\\
24,45.307042015196444,43.640893470196644,44.380770344636524\\
32,66.06244024954047,50.281284522797776,49.162105058614536\\
48,63.99956228781187,50.05435580260274,48.81611749269989\\
64,66.85390773645142,49.08609383555707,47.66504717130042\\
};
\nextgroupplot[xmin=5.76,xmax=66.24,ymin=0,ymax=80,ytick={0,20,40,60,80},xtick={8,16,32,48,64},xlabel={Active sessions}]
\addplot[forget plot,e2e background] table[x=level,y=background_ms,col sep=comma,row sep=\\] {
level,background_ms,hicache_ccoff_ms,velocc_ms\\
8,25.075667534677134,23.48078676390375,23.820597601413667\\
16,35.84939385557611,32.5798535978263,33.71880838798575\\
24,45.63570060585825,41.71069986508349,43.54867458945116\\
32,57.861963116171275,49.715629138228195,51.50180006049363\\
48,63.156060132815604,51.05817799338179,52.468375757440214\\
64,60.056567083286396,49.25209320345726,50.154350639459324\\
};
\addplot[forget plot,e2e cc off] table[x=level,y=hicache_ccoff_ms,col sep=comma,row sep=\\] {
level,background_ms,hicache_ccoff_ms,velocc_ms\\
8,25.075667534677134,23.48078676390375,23.820597601413667\\
16,35.84939385557611,32.5798535978263,33.71880838798575\\
24,45.63570060585825,41.71069986508349,43.54867458945116\\
32,57.861963116171275,49.715629138228195,51.50180006049363\\
48,63.156060132815604,51.05817799338179,52.468375757440214\\
64,60.056567083286396,49.25209320345726,50.154350639459324\\
};
\addplot[forget plot,e2e velocc] table[x=level,y=velocc_ms,col sep=comma,row sep=\\] {
level,background_ms,hicache_ccoff_ms,velocc_ms\\
8,25.075667534677134,23.48078676390375,23.820597601413667\\
16,35.84939385557611,32.5798535978263,33.71880838798575\\
24,45.63570060585825,41.71069986508349,43.54867458945116\\
32,57.861963116171275,49.715629138228195,51.50180006049363\\
48,63.156060132815604,51.05817799338179,52.468375757440214\\
64,60.056567083286396,49.25209320345726,50.154350639459324\\
};
\nextgroupplot[xmin=5.76,xmax=66.24,ymin=0,ymax=80,ytick={0,20,40,60,80},xtick={8,16,32,48,64},xlabel={Active sessions}]
\addplot[forget plot,e2e background] table[x=level,y=background_ms,col sep=comma,row sep=\\] {
level,background_ms,hicache_ccoff_ms,velocc_ms\\
8,27.269244901277336,26.3020025124574,26.489638648536875\\
12,36.88808556506864,35.386783482250614,36.106013947087554\\
16,46.13665463510783,43.25068948419335,43.576624606388215\\
24,60.3531851260132,54.375878675168934,51.67824178650355\\
32,58.47865536038558,55.37216933781903,51.302553398548\\
48,59.662042448409025,52.36414990265819,48.21317060440011\\
64,62.89113538721143,53.84097784104973,49.528943444306556\\
};
\addplot[forget plot,e2e cc off] table[x=level,y=hicache_ccoff_ms,col sep=comma,row sep=\\] {
level,background_ms,hicache_ccoff_ms,velocc_ms\\
8,27.269244901277336,26.3020025124574,26.489638648536875\\
12,36.88808556506864,35.386783482250614,36.106013947087554\\
16,46.13665463510783,43.25068948419335,43.576624606388215\\
24,60.3531851260132,54.375878675168934,51.67824178650355\\
32,58.47865536038558,55.37216933781903,51.302553398548\\
48,59.662042448409025,52.36414990265819,48.21317060440011\\
64,62.89113538721143,53.84097784104973,49.528943444306556\\
};
\addplot[forget plot,e2e velocc] table[x=level,y=velocc_ms,col sep=comma,row sep=\\] {
level,background_ms,hicache_ccoff_ms,velocc_ms\\
8,27.269244901277336,26.3020025124574,26.489638648536875\\
12,36.88808556506864,35.386783482250614,36.106013947087554\\
16,46.13665463510783,43.25068948419335,43.576624606388215\\
24,60.3531851260132,54.375878675168934,51.67824178650355\\
32,58.47865536038558,55.37216933781903,51.302553398548\\
48,59.662042448409025,52.36414990265819,48.21317060440011\\
64,62.89113538721143,53.84097784104973,49.528943444306556\\
};
\end{groupplot}
\coordinate (itlLegend) at ([yshift=9mm]group c2r1.north);
\begin{axis}[hide axis,at={(itlLegend)},anchor=south west,width=1cm,height=1cm,scale only axis,
xmin=0,xmax=1,ymin=0,ymax=1,legend columns=3,
legend style={at={(axis cs:0,0)},anchor=south,draw=none,fill=none,
font=\figurefont\fontsize{9}{10}\selectfont,cells={anchor=west},/tikz/every even column/.append style={column sep=8pt}}]
\addlegendimage{e2e background legend}\addlegendentry{\baseline (CC on)}
\addlegendimage{e2e velocc legend}\addlegendentry{\sysname{} (CC on)}
\addlegendimage{e2e cc off legend}\addlegendentry{HiCache (CC off)}
\end{axis}
\node[rotate=90,font=\figurefont\bfseries\fontsize{10}{11}\selectfont] at ([xshift=-17mm]group c1r1.west) {LooGLE};
\node[rotate=90,font=\figurefont\bfseries\fontsize{10}{11}\selectfont] at ([xshift=-17mm]group c1r2.west) {NarrativeQA};
\node[rotate=90,font=\figurefont\bfseries\fontsize{10}{11}\selectfont] at ([xshift=-17mm]group c1r3.west) {SWE-bench};
\end{tikzpicture}%
}

%% file: figures/gap-model-h200.tex
\begin{figure}[t]
\centering
\input{figures/gap-model-h200-plot}
\caption{Output-token throughput relative to CC-off for all 63 H200 configurations. Colors identify workloads, and shapes identify models. The line denotes equality and shading marks deviations of $\pm3$ percentage points.}
\label{fig:gap-model-h200}
\end{figure}
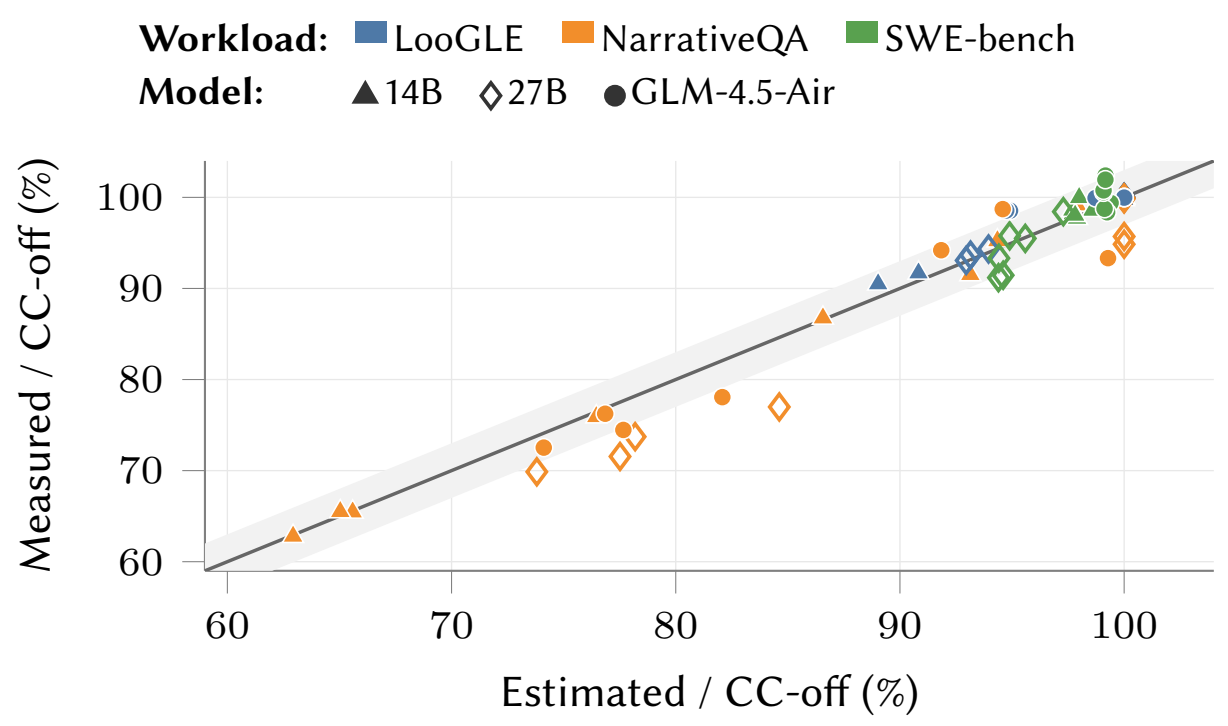

%% file: figures/gap-model-h200-plot.tex
\begingroup
\figurefont\selectfont
\begin{tikzpicture}
\begin{axis}[
    width=\columnwidth,
    height=0.52\columnwidth,
    xmin=59, xmax=104, ymin=59, ymax=104,
    xtick={60,70,80,90,100}, ytick={60,70,80,90,100},
    xlabel={Estimated / CC-off (\%)},
    ylabel={Measured / CC-off (\%)},
    label style={font=\figurefont\fontsize{8.5}{10}\selectfont},
    tick label style={font=\figurefont\fontsize{7.5}{9}\selectfont},
    tick align=outside, tick style={black!50},
    axis lines*=left, axis line style={black!50},
    grid=major, grid style={black!9, line width=.25pt},
    clip marker paths=true,
]
\addplot[draw=none,fill=black!5,forget plot]
    coordinates {(59,62) (101,104) (104,104) (104,101) (62,59) (59,59)} \closedcycle;
\addplot[black!60,line width=.65pt,forget plot] coordinates {(59,59) (104,104)};
\newcommand{\gapseries}[3]{%
    \addplot[only marks,#2,#3,forget plot,
        restrict expr to domain={\thisrow{series}}{#1:#1}]
        table[x=estimated_pct,y=measured_pct,col sep=comma,row sep=\\] {
series,model,workload,level,estimated_pct,measured_pct,estimate_without_gap_pct\\
0,14B,loogle,2,100.00000000,99.97802398,100.00000000\\
0,14B,loogle,3,100.00000000,99.96706819,100.00000000\\
0,14B,loogle,4,100.00000000,99.95587633,100.00000000\\
0,14B,loogle,5,100.00000000,99.75781754,100.00000000\\
0,14B,loogle,6,99.17095334,99.52373271,100.00000000\\
0,14B,loogle,8,90.82813842,91.69349291,92.48579387\\
0,14B,loogle,12,89.02713347,90.49562534,90.56733144\\
1,14B,narrativeqa,4,97.89978411,99.15878344,99.09890095\\
1,14B,narrativeqa,8,94.34821852,95.22703119,95.70415822\\
1,14B,narrativeqa,12,93.14003758,91.51682296,94.66245455\\
1,14B,narrativeqa,16,86.56540143,86.74996880,88.23938019\\
1,14B,narrativeqa,20,76.45689456,75.91306208,78.14433250\\
1,14B,narrativeqa,32,65.58832291,65.44852420,67.34913439\\
1,14B,narrativeqa,64,62.93214539,62.78508658,64.64277733\\
1,14B,narrativeqa,96,65.03616169,65.49345040,66.77134295\\
2,14B,swe,8,98.58185365,98.56868095,98.58185365\\
2,14B,swe,16,97.89949758,97.69008819,97.89949758\\
2,14B,swe,24,97.71199872,98.18473976,97.71199872\\
2,14B,swe,32,97.69028197,98.58456500,97.69028197\\
2,14B,swe,48,97.81570389,98.02309706,97.81570389\\
2,14B,swe,64,97.99351711,99.96317945,97.99351711\\
3,27B,loogle,2,100.00000000,99.99011944,100.00000000\\
3,27B,loogle,3,100.00000000,99.98116699,100.00000000\\
3,27B,loogle,4,100.00000000,99.96423822,100.00000000\\
3,27B,loogle,5,100.00000000,99.90940998,100.00000000\\
3,27B,loogle,6,93.95169169,94.33171361,93.95169169\\
3,27B,loogle,7,93.14874154,93.66276984,93.14874154\\
3,27B,loogle,8,92.95341223,93.07562237,92.95341223\\
4,27B,narrativeqa,4,100.00000000,99.94165685,100.00000000\\
4,27B,narrativeqa,8,100.00000000,99.83434555,100.00000000\\
4,27B,narrativeqa,16,100.00000000,95.70253021,100.00000000\\
4,27B,narrativeqa,24,100.00000000,94.85490862,100.00000000\\
4,27B,narrativeqa,36,84.61787638,76.99163275,91.66855363\\
4,27B,narrativeqa,48,78.18860867,73.72484475,88.86344343\\
4,27B,narrativeqa,64,77.51410824,71.55299383,85.17480798\\
4,27B,narrativeqa,96,73.80280029,69.86957899,83.40420029\\
5,27B,swe,8,97.28818869,98.42964087,97.28818869\\
5,27B,swe,16,95.59033294,95.49427488,95.59033294\\
5,27B,swe,24,94.89174920,95.77299588,94.89174920\\
5,27B,swe,32,94.41192110,93.32179855,94.41192110\\
5,27B,swe,48,94.39450089,91.18275387,94.39450089\\
5,27B,swe,64,94.60396862,91.44937720,94.60396862\\
6,GLM,loogle,1,100.00000000,99.99375773,100.00000000\\
6,GLM,loogle,1.50000000,100.00000000,99.98845753,100.00000000\\
6,GLM,loogle,2,100.00000000,99.97723205,100.00000000\\
6,GLM,loogle,3,100.00000000,99.97219969,100.00000000\\
6,GLM,loogle,4,98.72620166,99.93606541,100.00000000\\
6,GLM,loogle,6,94.91014471,98.53654203,98.06949118\\
6,GLM,loogle,8,94.74004570,98.55686546,97.99746629\\
7,GLM,narrativeqa,4,94.58447859,98.71071407,95.54045390\\
7,GLM,narrativeqa,8,91.84223672,94.19691440,93.30551378\\
7,GLM,narrativeqa,12,99.28042558,93.32783665,100.00000000\\
7,GLM,narrativeqa,16,82.07726114,78.06750585,85.29160299\\
7,GLM,narrativeqa,20,76.85261073,76.24543862,80.34486695\\
7,GLM,narrativeqa,24,77.66279593,74.46771262,81.33000241\\
7,GLM,narrativeqa,32,74.11241951,72.52601970,77.84290237\\
8,GLM,swe,8,99.39542942,99.39274615,99.39542942\\
8,GLM,swe,12,99.24134044,98.36320176,99.24134044\\
8,GLM,swe,16,99.12512755,98.75173685,99.12512755\\
8,GLM,swe,24,99.08101881,100.56035690,99.08101881\\
8,GLM,swe,32,99.08101881,100.76843143,99.08101881\\
8,GLM,swe,48,99.16831542,102.39063244,99.16831542\\
8,GLM,swe,64,99.16351491,101.94079506,99.16351491\\
};%
}
\gapseries{0}{VeloCCProtected}{mark=triangle*,mark size=2.3pt,mark options={draw=white,line width=.3pt}}
\gapseries{1}{VeloCCCipher}{mark=triangle*,mark size=2.3pt,mark options={draw=white,line width=.3pt}}
\gapseries{2}{VeloCCBackup}{mark=triangle*,mark size=2.3pt,mark options={draw=white,line width=.3pt}}
\gapseries{3}{VeloCCProtected}{mark=diamond,mark size=2.4pt,mark options={line width=.7pt}}
\gapseries{4}{VeloCCCipher}{mark=diamond,mark size=2.4pt,mark options={line width=.7pt}}
\gapseries{5}{VeloCCBackup}{mark=diamond,mark size=2.4pt,mark options={line width=.7pt}}
\gapseries{6}{VeloCCProtected}{mark=*,mark size=1.7pt,mark options={draw=white,line width=.3pt}}
\gapseries{7}{VeloCCCipher}{mark=*,mark size=1.7pt,mark options={draw=white,line width=.3pt}}
\gapseries{8}{VeloCCBackup}{mark=*,mark size=1.7pt,mark options={draw=white,line width=.3pt}}
\end{axis}
\newcommand{\gaplegendcolor}[1]{%
    \tikz[baseline=-.5ex]\fill[#1] (0,0) rectangle (6pt,4pt);%
}
\newcommand{\gaplegendmark}[2]{%
    \tikz[baseline=-.5ex]\draw[VeloCCInk,mark=#1,mark size=#2,line width=.7pt]
        plot coordinates {(0,0)};%
}
\node[anchor=south,inner sep=0pt,outer sep=0pt,
    font=\figurefont\fontsize{7.5}{9}\selectfont]
    at ([yshift=4pt]current bounding box.north) {%
    \setlength{\tabcolsep}{0pt}%
    \begin{tabular}{@{}l@{\hspace{5pt}}l@{}}
    \textbf{Workload:} &
        \gaplegendcolor{VeloCCProtected}\,LooGLE\hspace{7pt}%
        \gaplegendcolor{VeloCCCipher}\,NarrativeQA\hspace{7pt}%
        \gaplegendcolor{VeloCCBackup}\,SWE-bench\\[1pt]
    \textbf{Model:} &
        \gaplegendmark{triangle*}{2.3pt}\,14B\hspace{7pt}%
        \gaplegendmark{diamond}{2.4pt}\,27B\hspace{7pt}%
        \gaplegendmark{*}{1.7pt}\,GLM-4.5-Air
    \end{tabular}%
};
\end{tikzpicture}
\endgroup